%% file: main_arxiv.tex
\documentclass[11pt]{article}
\usepackage{amssymb,amsbsy,amsfonts,amsmath,amsthm,xspace}
\usepackage{mathrsfs}
\usepackage{graphicx}
\usepackage{caption}
\usepackage{setspace}
\usepackage{enumitem}
\usepackage{subcaption}
\usepackage{multirow}
\usepackage{epstopdf}
\usepackage{epsfig}
\usepackage{url}
\usepackage[toc,page]{appendix}
\usepackage{float}
\usepackage{natbib}
\usepackage{color}
\usepackage{bm}
\usepackage{verbatim}
\usepackage{authblk}
\usepackage[normalem]{ulem}
\usepackage{hyperref}
\hypersetup{
    colorlinks=true,
    linkcolor=blue,
    filecolor=magenta,
    urlcolor=cyan,
    citecolor=blue
}

\usepackage{times}

\usepackage{subfiles}
\graphicspath{{art/}}

\theoremstyle{definition}
\newtheorem{theorem}{Theorem}

\newtheorem{proposition}{Proposition}

\newtheorem{lemma}{Lemma}
\newtheorem{remark}{Remark}
\newtheorem{example}{Example}
\newtheorem{assumption}{Assumption}

\usepackage[plain,noend]{algorithm2e}

\newcommand{\tp}{\intercal}
\newcommand{\expect}{\mathbb{E}}
\newcommand{\prob}{\mathbb{P}}
\newcommand{\hatbar}[1]{\hat{\bar{ #1 }}}
\newcommand{\neweps}{\varepsilon}
\newcommand{\diag}{\operatorname{diag}}
\newcommand{\rank}{\operatorname{rank}}
\newcommand{\col}{\operatorname{col}}
\newcommand{\row}{\operatorname{row}}
\newcommand{\iprod}[2]{\langle #1 , #2 \rangle}

\newcommand{\subdiff}[1]{\partial \lVert #1 \rVert_*}
\newcommand{\termi}{(\mathcal{I})}
\newcommand{\termii}{(\mathcal{I}\mathcal{I})}
\newcommand{\termiii}{(\mathcal{I}\mathcal{I}\mathcal{I})}
\newif\ifbka
\bkafalse

\begin{document}

\renewcommand{\baselinestretch}{1}

\title{Latent space models for multiplex networks with shared structure}

\author{Peter W. MacDonald, Elizaveta Levina, and Ji Zhu \\ Department of Statistics, University of Michigan}
\date{July 7, 2021}

\maketitle

\begin{abstract}
  Latent space models are frequently used for modeling single-layer networks and include many popular special cases, such as the stochastic block model and the random dot product graph. However, they are not well-developed for more complex network structures, which are becoming increasingly common in practice. Here we propose a new latent space model for multiplex networks: multiple, heterogeneous networks observed on a shared node set. Multiplex networks can represent a network sample with shared node labels, a network evolving over time, or a network with multiple types of edges. The key feature of our model is that it learns from data how much of the network structure is shared between layers and pools information across layers as appropriate. We establish identifiability, develop a fitting procedure using convex optimization in combination with a nuclear norm penalty, and prove a guarantee of recovery for the latent positions as long as there is sufficient separation between the shared and the individual latent subspaces. We compare the model to competing methods in the literature on simulated networks and on a multiplex network describing the worldwide trade of agricultural products.
\end{abstract}


\subfile{intro}

\subfile{model}

\subfile{estimation}

\subfile{theory}

\subfile{simulations}

\subfile{real_data}

\subfile{conclusion}


\bibliographystyle{abbrvnat}
\bibliography{mybib}

\pagebreak

\appendix



\subfile{appendix_optimization}

\subfile{appendix_proofs}

\subfile{appendix_sims}

\end{document}

%% file: intro.tex
\section{Introduction}


Network data have become commonplace in many statistical applications, including neuroscience, social sciences, and computational biology, among others.
In the vast majority of cases, these network data are represented as {\em graphs}.
At a minimum, a graph $G=(V,E)$ has a node set $V$ and an edge set $E$, with each edge connecting a pair of nodes, but frequently additional information is available, such as node attributes, edge weights, multiple types of edges, and so on.
While a lot of work has been done on a single network with binary edges, as the complexity of network data structure increases, the availability of statistical methods and models dwindles rapidly.
There is a strong need for rigorous statistical analysis to keep up with the rapidly increasing complexity of real datasets.

One such complex network data structure is the {\em multilayer graph} \citep{kivela14multilayer}, a highly general mathematical object which can describe multiple graphs, dynamic graphs, hypergraphs, and vertex-colored or edge-colored graphs.
In addition to a node set and an edge set, a multilayer graph includes a layer set. A node may appear on any or all layers, and each edge connects two vertices, including the possibility of connection in the same layer, an {\em intra-layer edge}; across layers, an {\em inter-layer edge}; and between the same node on different layers.
For example, a general multilayer network could be used to represent a multi-modal urban transportation network of bus, train, bicycle, and other connections, where each layer corresponds to a different mode of transportation, and edges define connections between stations.

The focus of this paper is on multiplex graphs, a type of multilayer graph where a common set of $n$ nodes appears on every layer, and no inter-layer edges are allowed.
For example, the brain connectivity networks of a sample of people, or a multi-commodity international trade network
could be represented as a multiplex network where each layer corresponds to a subject, or a commodity respectively.


For the single undirected graph $G$ with $\lvert V(G)\rvert = n$, a common approach to modeling is to assume that there are $n$ latent variables $\{X_i\}_{i=1}^n \subseteq \mathcal{X}$, one for each node.
Typically one further assumes that for each node pair $i \leq j$, $X_i$ and $X_j$ fully parameterize the distribution of the edge variable $E_{ij} = \mathbf{1}\left\{ (i,j) \in E(G) \right\}$, and all the edge variables are mutually independent \citep{matias14modeling}.
The latent positions themselves are sometimes treated as fixed and sometimes as independent random variables;  in the latter case the above assumptions are conditional on $\{X_i\}_{i=1}^n$.
These models are called {\em latent space} models, and intuitively the latent variable $X_i$ represents the behavior of node $i$ through its position in the latent space $\mathcal{X}$.
\cite{matias14modeling} distinguish between two cases: a discrete latent space $\mathcal{X} = \{1,\ldots, K\}$, so that each node is in one of $K$ latent classes, and  $\mathcal{X} = \mathbb{R}^d$, so that each node is represented by its coordinates in Euclidean space.
The first case includes the ubiquitous {\em stochastic block model} (SBM) \citep{holland83stochastic}, and a well-studied example of the second case is the {\em random dot product graph} (RDPG)  \citep{athreya17statistical,young07random}.
Models in the seminal papers of \cite{hoff02latent} and \cite{handcock07model} correspond to the second case as well.


Some of the frequentist approaches to latent space models treat the latent variables as random and focus on estimation of and inference for the parameters governing their distribution(s), for example \cite{bickel13asymptotic} in the SBM setting.
Many others do inference conditional on the latent variables and estimate them, especially when the goal is community detection, for example, \cite{lei15consistency} in the SBM setting, \cite{athreya17statistical} in the RDPG setting,  and \cite{ma20universal} in a latent space model with edge covariates.


Some extensions of latent space models for multilayer networks have been proposed in the literature.
This can be divided into two general categories: general multiple networks, for instance from repeated measurements or multiple subjects; and dynamic or time-varying networks, for which the layers have a natural ordering.
A review paper \citep{kim18review} details recent developments in the dynamic setting.

In the multiple networks setting, latent space models with a Bayesian approach to estimation have been proposed by \cite{gollini16joint}, \cite{salter17latent}, \cite{dangelo19latent}, and \cite{sosa21latent}, among others.
While hierarchical Bayesian approaches allow these models to adaptively share information or model dependence across layers, they tend to be computationally expensive for large networks.

For larger networks we aim to work with, we will focus on three recent frequentist approaches to latent space and low-rank modeling for multiple networks, as baselines to contrast to our proposal in this paper.
\cite{arroyo19inference} consider a collection of independent RDPGs with a common invariant subspace.
That is, the expected adjacency matrices for each layer are assumed to share a common, low-dimensional column space.
This is similar to approaches taken by \cite{levin17central}, \cite{wang19joint}, \cite{nielsen18multiple}, and \cite{jones20multilayer}.
However, they do not consider the case where each layer also contains meaningful individual signal in addition to shared structure.

\cite{zhang20flexible} consider a model where expected adjacency matrices, after a logistic transformation, share common low-rank structure. This framework allows for layer-specific parameters controlling degree heterogeneity, but no other individual structure.

\cite{wang19common} aim to decompose each expected adjacency matrix into a common and individual part after applying a logistic transformation.
They assume that the individual part is low-rank, but make no such assumptions on the common part.
Thus this method loses the interpretability afforded by the latent space approach, and has high variability unless there are a large number of layers.

Finally, our model bears a resemblance to other recent work which aims to summarize  multiple matrix-valued observations outside of the networks setting.   For example, \cite{lock20bidimensional}, in the setting of {\em multiview} data, propose a joint and individual approach to matrix factorization. \cite{devito19multi} propose a model for multi-study factor analysis which estimates both common and individual factors.

In extending latent space models to the multiple networks setting, we seek a modeling approach which can leverage shared structure to improve estimation accuracy, but in an adaptive way, learning how much the layers have in common from the data instead of assuming that the entire latent representation is shared across all layers.
We also allow for non-trivial individual structure in order to robustly estimate truly common structure.

As a motivating example, which we will analyze in Section~\ref{section_realdata}, consider
a multiplex network of international trade, where nodes correspond to countries, layers to different commodities, and each weighted intra-layer edge is the total trade of a given agricultural commodity between two nations.
We would expect the structure in this network to be governed by node attributes corresponding to, for instance, geographical region, language or climate.
Some of these attributes would be expected to affect all commodities similarly;  geographical proximity would encourage trade of any commodity.
On the other hand, some of these attributes may differ across layers; climate may encourage production and thus trade of some commodities but not others, depending on what crops are easiest to grow in a given country's climate.
In a setting like this, if latent space models were fit to each layer individually, (1) a fitting procedure cannot leverage the shared structure across layers, and (2) the latent representation of the common structure will not be automatically aligned across layers.
On the other hand, if a single latent space model is fit to all network layers jointly, or to some aggregated version, (1) an influential individual latent dimension, or one that is shared by some but not all layers, may be erroneously identified as a common effect; or (2) the influence of a common latent dimension may be overstated if it is not orthogonal to the individual latent dimensions.  The model we propose in the next section aims to address these shortcomings.



%% file: model.tex
\section{A new model for multiplex networks}
\label{section_model}

\subsection{Multiplex networks with shared structure}

Here we propose a new model for MULTIplex NEtworks with Shared Structure (MultiNeSS) 
with the goal of ultimately learning the amount of shared structure from data.
We start by fixing notation.
Suppose that we observe $m$ undirected networks, weighted or unweighted, on a common set of $n$ nodes with no self loops.
The networks are represented by their $n \times n$ adjacency matrices $\{A_k\}_{k=1}^m$.
Each node $i$ is associated with a fixed latent position describing its function in layer $k$, denoted $x_{k,i} \in  \mathbb{R}^d$.  The edges are assumed independent conditional on these latent positions:
\begin{equation*}
A_{k,ij} \overset{\text{ind}}{\sim} Q(\cdot;\kappa(x_{k,i},x_{k,j}),\phi) \quad (i=1,\ldots,n;\ j=1,\ldots, n;\ i < j; \ k=1,\ldots, n ),
\end{equation*}
where $Q(\cdot;\theta,\phi)$ is some edge entry distribution with a scalar parameter $\theta$ and possible nuisance parameters $\phi$, and $\kappa(\cdot,\cdot)$ is a symmetric similarity function, implying that the parameter $\theta$ captures the effect of the latent similarity of nodes $i$ and $j$ in layer $k$ on the corresponding edge.
We denote the latent positions for layer $k$ by $X_k$, where the $i$th row of the $n \times d_k$ matrix $X_k$ corresponds to the latent position of node $i$ in layer $k$. In general, the position as well as its dimension may depend on the layer $k$.

The choice of similarity function $\kappa$ may affect identifiability of each $X_k$.
For instance, if $\kappa(x,y) = \psi(x^{\tp}y)$ is an invertible scalar function $\psi$ applied to the Euclidean inner product, each $X_k$ is only identifiable up to a common orthogonal rotation of the rows.
If $\kappa(x,y) = \psi(||x-y||_2)$ is similarly defined as an invertible function of the Euclidean distance rather than the Euclidean inner product, $X_k$ is only identifiable up to a common orthogonal rotation and/or reflection of the rows, and a common shift of each row by a vector in $\mathbb{R}^{d_k}$. 

The key assumption of the MultiNeSS model is that some, but not all, structure is shared across network layers.
We suppose that the matrix $X_k$ can be written as
\begin{equation}
  X_k = \begin{bmatrix} V & U_k \end{bmatrix} \quad (k=1,\ldots, m),
  \label{eq:modelX}
\end{equation}
where $ V \in \mathbb{R}^{n \times d_1}$ gives a matrix of {\em common} latent position coordinates, and $U_k \in \mathbb{R}^{n \times d_{2,k}}$ are {\em individual} latent position coordinates for layer $k$.
Writing $X_k$ in this way further complicates identifiability.
The model will certainly be identifiable only up to some invariant transformation of the rows of each $U_k$, and of $V$, but we would still want $V$ to be identifiable in such a way that it is aligned across all the layers.
Intuitively, for this to hold we need the common dimension $d_1$ to be maximal and unique, in the sense that any transformation which aligns the first $d_1$ coordinates must partition $X_k$ into $V$ and $U_k$ as written above.
We will formalize this intuition in Section~\ref{id}.
First, we present some concrete examples of latent space models which fit the general MultiNeSS model framework.



\begin{example}[Low rank, Gaussian errors] \label{model1}
As a simple example, let the similarity function for each layer be the generalized inner product as described in \cite{rubin17statistical}. For vectors $x$ and $y$ in $\mathbb{R}^{p+q}$,
\begin{equation*}
  \kappa_{p,q}(x,y) = x_1y_1 + \cdots + x_py_p - x_{p+1}y_{p+1} - \cdots - x_{p+q}y_{p+q} = x^{\tp} I_{p,q} y
\end{equation*}
where $I_{p,q}$ is a block diagonal matrix
\begin{equation*}
  I_{p,q} = \begin{bmatrix}
    I_p & 0 \\ 0 & -I_q
\end{bmatrix},
\end{equation*}
and throughout the paper $I_r$ for a positive integer $r$ denotes the $r \times r$ identity matrix.
Under the generalized inner product, the first $p$ latent dimensions are referred to as {\em assortative}, while the remaining $q$ are {\em disassortative} \citep{rubin17statistical}.

Assume $Q(\cdot;\theta,\sigma)$ is the Gaussian distribution $\mathcal{N}(\theta,\sigma^2)$.
Then each layer's adjacency matrix has expectation
\begin{equation*}
\expect(A_k) = P_k = VI_{p_1,q_1}V^{\tp} + U_k I_{p_{2,k},q_{2,k}} U_k^{\tp} \quad (k=1,\ldots , m), 
\end{equation*}
where $p_1 + q_1 = d_1$, $p_{2,k} + q_{2,k} = d_{2,k}$, and
Each error matrix $E_k = A_k - \expect(A_k)$ is symmetric with i.i.d.\ mean 0 Gaussian entries.
\end{example}

If the setting does not allow for self-loops, we can instead use
\begin{equation*}
  \expect(A_k) = P_k = VI_{p_1,q_1}V^{\tp} + U_k I_{p_{2,k},q_{2,k}} U_k^{\tp}  - \diag (VI_{p_1,q_1} V^{\tp} + U_kI_{p_{2,k},q_{2,k}} U_k^{\tp}) \quad (k=1,\ldots , m)
\end{equation*}
to enforce zeros on the diagonal.
The same can be done in any of the subsequent examples, if needed.


\begin{example}[Low rank, exponential family errors] \label{model2}  Let the similarity function $\kappa$ be the generalized inner product again, and let $Q(\cdot;\theta)$be a one-parameter exponential family distribution with natural parameter $\theta$ and log-partition function $\nu$.
That is,
\begin{equation*}
Q(x;\theta) \propto \exp \left\{ x \theta - \nu(\theta) \right\}.
\end{equation*}
For instance, $Q$ may be a Bernoulli distribution, in which case $\nu(\theta) = \log(1+e^{\theta})$.
In the spirit of generalized linear models, we model the edges by applying the canonical link function $g = \nu'$ entry-wise, so that adjacency matrices now satisfy
\begin{equation*}
\expect(A_{k,ij}) = P_{k,ij} = g(v_i^{\tp}I_{p_1,q_1} v_j + u_{k,i}^{\tp}I_{p_{2,k},q_{2,k}} u_{k,j}) \quad (i=1,\ldots,n;\ j=1,\ldots, n;\ i \leq j; \ k=1,\ldots, n ).
\end{equation*}
In the Bernoulli example, the canonical link function is the inverse logistic function
\begin{equation} \label{expit}
  g(\theta) = \frac{e^{\theta}}{1+e^{\theta}}.
\end{equation}
\end{example}


\subsection{Identifiability} \label{id}

We present a sufficient condition for identifiability when $\kappa$ is a scalar function of the generalized inner product. For more detailed discussion of the statistical implications of such transformations, see \cite{rubin17statistical}.
For other choices of similarity function, conditions for identifiability will depend on the set of invariant transformations which it induces.

For the inner product model with one layer, it is natural to assume that the matrix of latent positions $X \in \mathbb{R}^{n \times d}$ is full rank, that is, it has linearly independent columns.
We show that a stronger linear independence condition for all pairwise concatenations of the latent position matrices is sufficient for identifiability in the proposed MultiNeSS model.
The proof is given in Appendix~\ref{appendix_prop1}.


\begin{proposition} \label{identifiability}
Suppose $\kappa_{p,q}(x, y) = \psi(x^{\tp}I_{p,q} y)$ is an invertible scalar function of the generalized inner product, and
the model is parameterized by $V$ and $\{U_k\}_{k=1}^m$ as in \eqref{eq:modelX}.

Define an undirected graph $\mathcal{G}_{I}$ on the network {\em layers},
with vertex set $\{1,\ldots ,m\}$, and edges
\begin{equation} \label{lin_independence}
k \sim l \iff \begin{bmatrix} V & U_{k} & U_{l} \end{bmatrix} \quad \text{is linearly independent.}
\end{equation}
If $\mathcal{G}_{I}$ is connected, then the model is identifiable up to indefinite orthogonal transformation. That is, if the probability distributions induced by two different parameterizations $(V,U_1,\ldots ,U_m)$ and  $(V',U'_1,\ldots ,U'_m)$ coincide,  then
\begin{equation*}
V = V'W_0,U_1 = U'_1W_1,\ldots ,U_m=U'_mW_m
\end{equation*}
for some indefinite orthogonal transformations $\{W_k\}_{k=0}^m$.
\end{proposition}

To simplify the condition in Propostion~\ref{identifiability}, consider the special case in which $\mathcal{G}_I$ is the complete graph, equivalent to assuming that for all $1 \leq k_1 < k_2 \leq m$, the $n \times (d_1 + d_{2,k_1} + d_{2,k_2})$ matrix
\begin{equation*} 
\begin{bmatrix} V & U_{k_1} & U_{k_2} \end{bmatrix}
\end{equation*}
has linearly independent columns. In the special case when $q_1 = q_{2,1} = \cdots = q_{2,m} = 0$, the similarity function for the latent vectors is standard Euclidean inner product, and Proposition~\ref{identifiability} holds with identifiability up to orthogonal rotation.

If we assume that the fully concatenated $n \times (d_1 + \sum_k d_{2,k})$ matrix
\begin{equation*} 
\begin{bmatrix} V & U_{1} & \cdots & U_{m} \end{bmatrix}
\end{equation*}
has linearly independent columns, as in \cite{devito19multi} for a similar factor analysis model, then once again $\mathcal{G}_I$ will be the complete graph and Proposition~\ref{identifiability} will hold. Note that Proposition~\ref{identifiability} does not require the orthogonality of the columns of $V$ and $\{U_k\}_{k=1}^m$, although clearly it will be satisfied if the columns are all mutually orthogonal.

As a visual intuition, consider a simple case with $n=10$, $d_1=d_2=1$, $m=2$. Standard results for the RDPG \citep{athreya17statistical} would suggest that the two-dimensional latent positions for each layer are only identifiable up to orthogonal rotations, which differ across layers. The recovered latent positions for the two layers may have different rotations, and thus would not share a common column according to the MultiNeSS model \eqref{eq:modelX}. Proposition~\ref{identifiability} states that pairwise linear independence is sufficient to uniquely align the rotations, and identify the common and individual latent positions up to sign.

\begin{figure}
    \centering
    \includegraphics[width=.8\textwidth]{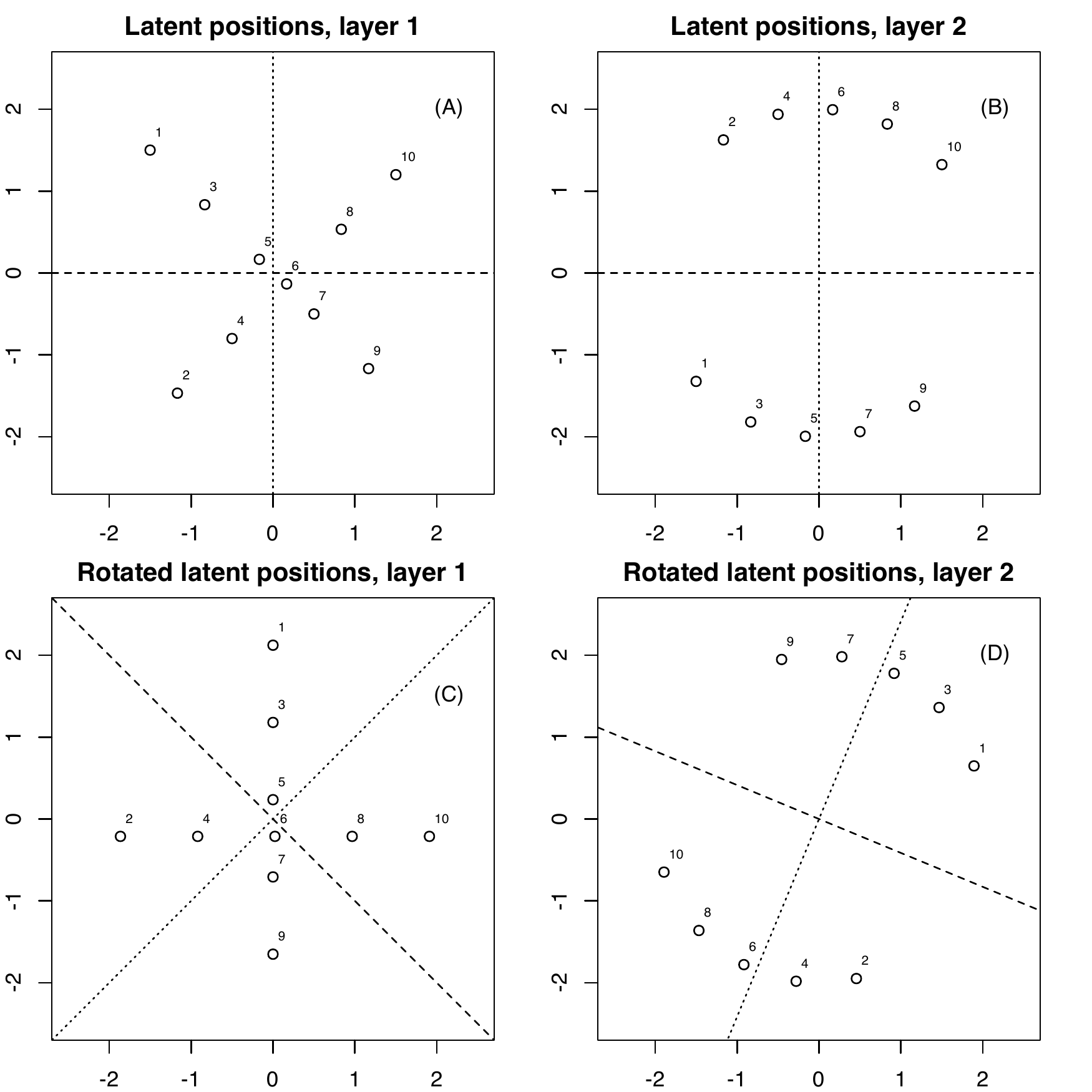}
    \caption{Latent positions before (top row) and after (bottom row) orthogonal rotation.}
    \label{identifiability_example}
\end{figure}

In Figure~\ref{identifiability_example} panels (A) and (B), we plot latent positions $\{(v_i,u_{k,i})\}_{i=1}^n$ in $\mathbb{R}^2$ for $k=1$ and $2$ respectively.   Each point is labeled with its index, 1 through 10, for ease of matching across the panels.
The common dimension is on the $x$-axis and the individual dimension on the $y$-axis; thus the $x$-coordinates are the same in panels (A) and (B).
In Figure~\ref{identifiability_example} panels (C) and (D), we apply an orthogonal transformation to each of the latent positions, equivalent to applying an unknown two-dimensional orthogonal rotation.
After rotation, the points in panel (C) do not match the points in panel (D) in either their $x$  or the $y$ coordinates.
The dashed lines denote the original $x$-axis in the two rotated spaces; note that the coordinates of projection onto these directions are constant in the top two panels.
After rotation, we identify directions, denoted by dotted lines in panels (C) and (D), with the property that for all points the coordinates of projection onto these directions are the same in the bottom two panels.
By Proposition~\ref{identifiability}, as long as \eqref{lin_independence} holds, the original $x$-axis is the unique direction with this property, and the coordinates of projection uniquely identify the entries of $v$, up to sign.

%% file: estimation.tex
\section{Fitting the MultiNeSS model} \label{methods}

\subsection{Convex objective function}

A natural approach to latent space estimation is likelihood maximization. A convex relaxation of the likelihood can be maximized by introducing a nuclear norm penalty, and by optimizing over the entries of the low-rank matrices $F = VI_{p_1,q_1}V^{\tp}$ and $G_k = U_kI_{p_{2,k},q_{2,k}}U_k^{\tp}$ for each $k=1,\ldots m$ rather than the latent position matrices themselves.

With the notation defined in Section~\ref{section_model}, suppose $\kappa$ is a generalized inner product on $\mathbb{R}^d$. In terms of the latent position parameters $(V,\{U_k\}_{k=1}^m)$, the negative log-likelihood, after dropping constants, takes the form
\begin{equation} \label{log_likelihood}
\ell \left( V,\{U_k\}_{k=1}^m \mid \{A_k\}_{k=1}^m \right) \propto - \sum_{k=1}^m \sum_{i \leq j} \log Q(A_{k,ij};v_i^{\tp}I_{p_1,q_1}v_j + u_{k,i}^{\tp}I_{p_{2,k},q_{2,k}}u_{k,j},\phi),
\end{equation}
where $Q$ is the density of the edge weight distribution.
Up to a rotation, we can rewrite this likelihood in terms of symmeric $n \times n$  matrices $F = VI_{p_1,q_1} V^{\tp}$ and $G_k = U_k I_{p_{2,k},q_{2,k}} U_k^{\tp} \ (k=1,\ldots m)$ by constraining the number of positive and negative eigenvalues of each matrix.
For a symmetric matrix $M$, let $\operatorname{r}^+(M)$ and $\operatorname{r}^-(M)$ respectively denote the number of strictly positive and strictly negative eigenvalues of $M$.
Then we equivalently minimize
\begin{equation*}
\ell \left( F,\{G_k\}_{k=1}^m \mid \{A_k\}_{k=1}^m \right) = - \sum_{k=1}^m \sum_{i \leq j} \log Q(A_{k,ij}; F_{ij} + G_{k,ij},\phi)
\end{equation*}
subject to the constraints
\begin{align*}
\operatorname{r}^+(F) &\leq p_1, \quad \operatorname{r}^+(G_k) \leq p_{2,k} \quad (k=1,\ldots m), \\
\operatorname{r}^-(F) &\leq q_1, \quad \operatorname{r}^-(G_k) \leq q_{2,k} \quad (k=1,\ldots m).
\end{align*}
When $\kappa$ is the Euclidean inner product, $q_1 = q_{2,1} = \cdots = q_{2,m} = 0$, and the constraints are equivalent to requiring each matrix to be low-rank and positive semi-definite.

To make this problem tractable, we ignore the constraint on the eigenvalue signs, and perform a further convex relaxation of the resulting
rank constraint, leading to the unconstrained convex optimization problem
\begin{equation} \label{ls_objective}
\underset{F,G_k}{\text{min}}  \left\{ \ell \left( F,\{G_k\}_{k=1}^m \mid \{A_k\}_{k=1}^m \right) + \lambda \lVert F \rVert_* + \sum_{k=1}^m \lambda \alpha_k \lVert G_k \rVert_* \right\},
\end{equation}
where $\lambda \geq 0$ and $\alpha_k \geq 0\ (k=1,...,m)$ are tuning parameters, and $\lVert \cdot \rVert_*$ denotes the nuclear norm of a matrix, equal to the sum of the singular values.
The parameter $\lambda$ appears in both terms as an overall scaling depending on $n$ and the total entry-wise variance across all the layers, while each $\alpha_k$ controls the individual penalties, depending on the entry-wise variance of each layer.
Since the nuclear norm is convex, it is easy to see that \eqref{ls_objective} defines a convex optimization problem as long as the edge distribution $Q$ is log-concave in $\theta$.




\subsection{Proximal gradient descent algorithm}

The optimization problem can be solved by applying proximal gradient descent block-wise to each of the matrix arguments. In particular, we split the optimization variables into $m+1$ blocks of $n^2$ variables: one block containing the entries of $F$; and $m$ blocks, one for the entries of each $G_k$.

Then for each block, the negative log-likelihood is convex and differentiable, and the nuclear norm penalty term is convex and although non-differentiable, it has a well-defined proximal mapping for step size $\eta >0$ \citep{fithian18flexible}. In particular, the nuclear norm scaled by $\lambda \geq 0$ has the proximal mapping
\begin{equation*}
\operatorname{argmin}_{M'} \frac{1}{2\eta} \lVert M - M' \rVert_F^2 + \lambda \lVert M' \rVert_* =S_{\eta\lambda}(M),
\end{equation*}
where $\lVert \cdot \rVert_F$ denotes the matrix Frobenius norm, the Euclidean norm of the vectorized entries; and $S_{T}(\cdot)$ is the soft singular value thresholding operator with threshold $T \geq 0$.
That is, for a diagonal matrix $M \in \mathbb{R}^{q \times q}$,
\begin{equation*}
  S_T(M) = \operatorname{diag}\{(M_{11}-T)_+,\ldots ,(M_{qq}-T)_+\},
\end{equation*}
and otherwise $S_T(M) = US_T(D)V^{\tp}$, where $M = UDV^{\tp}$ is the singular value decomposition (SVD) of $M$ \citep{fithian18flexible}.

Thus, we derive the following proximal gradient descent steps with step size $\eta/m$ for updates of $F$, and $\eta$ for updates of each $G_k$: at iteration step $t \geq 1$,
\begin{align*}
\hat{F}^{(t)} &= S_{\eta \lambda /m} \left\{ \hat{F}^{(t-1)} + \frac{\eta}{m}\frac{\partial}{\partial F} \ell(\hat{F}^{(t-1)},\{ \hat{G}_{k'}^{(t-1)}\}_{k'=1}^m ) \right\}, \\
\hat{G}_k^{(t)} &= S_{\eta\lambda \alpha_k} \left\{\hat{G}_k^{(t-1)} + \eta \frac{\partial}{\partial G_k} \ell(\hat{F}^{(t)},\{ \hat{G}_{k'}^{(t-1)}\}_{k'=1}^m)\right\} \quad (k=1,\ldots ,m).
\end{align*}
This particular choice of relative step sizes is discussed in more detail in Appendix~\ref{appendix_pgd}.


When $Q$ is a one-parameter exponential family, and the edge distribution is modeled through the canonical link function, see Example~\ref{model2}, the gradients take on a particularly nice form. In particular, up to an additive constant, for each fixed node pair $(i,j)$,
\begin{equation*}
(\log Q)'(A_{k,ij};F_{ij}+G_{k,ij}) = A_{k,ij} - \expect(A_{k,ij} ; F_{ij}+G_{k,ij}) \quad (k=1,\ldots m),
\end{equation*}
where the equality follows by the choice of link function $g = \nu'$. Thus the gradients with respect to each $G_k$ can be interpreted as the residual from estimating the adjacency matrix by its expectation given the current low-rank parameters. The gradient with respect to $F$ is the sum of the residuals over all the layers.

If $Q(\cdot;\theta,\phi)$ is the Gaussian distribution, as in Example~\ref{model1}, the appropriate link function is the identity link.
Then with step size $1/m$ for updates of $F$, and $1$ for updates of each $G_k$, proximal gradient descent recovers a natural alternating soft-thresholding algorithm: at iteration step $t \geq 1$,
\begin{equation} \label{update_theory}
\hat{F}^{(t)} = S_{\lambda /m} \left\{ \frac{1}{m}\sum_{k=1}^m (A_k - \hat{G}^{(t-1)}_k) \right\}, \quad \hat{G}^{(t)}_k = S_{\lambda\alpha_k}\left( A_k - \hat{F}^{(t)} \right)\ (k=1,\ldots ,m).
\end{equation}
In Section~\ref{section_theory}, we will provide theoretical guarantees on estimators found with this special case of proximal gradient descent.
When the observed networks have no self-loops, we will perform proximal gradient steps which ignore the diagonal entries, which should provide better empirical results in this case.

While we presented the algorithm for the generalized inner product similarity, it can easily be adapted to the usual inner product similarity by enforcing a positive semi-definite constraint on $F$ and on each $G_k$ in \eqref{ls_objective}.
The constraint can be enforced by adding a positive semi-definite projection step at each iteration of the proximal gradient descent algorithm, equivalent to shrinking the negative eigenvalues of each iterate to zero.

The most computationally expensive part of each update is the SVD needed for soft singular value thresholding. If the full SVD is calculated, each iteration step has computational complexity of order $O(mn^3)$. In practice,  we use a truncated SVD which only finds the first $s \ll n$ singular vectors and values, as in \cite{wu17generalized}, reducing complexity to $O(mn^2s)$. 
For synthetic multiplex networks generated according to the models from  Examples \ref{model1} and \ref{model2} with $n=400$ and $m=8$, see Section~\ref{section_simulations}, our R implementation of proximal gradient descent is able to perform approximately one iteration per second. When the signals are sufficiently strong, the algorithm typically converges in fewer than 10 steps.

\subsection{Refitting step} \label{subsection_refit}

As we will show in Theorem~\ref{thm1}, recovery of the correct rank requires a tuning parameter of order $\lambda \sim \sqrt{n}$,
and thus the effect of the soft thresholding step on the estimated eigenvalues will not disappear as $n \rightarrow \infty$.

As in \cite{mazumder10spectral}, we propose a refitting step after solving the convex problem, where we fix the ranks and eigenvectors of the estimated $\hat{F}$ and $\hat{G}_k$, and refit their eigenvalues to maximize the original non-convex likelihood.



Based on the output from the first step, we write the eigendecompositions
\begin{equation} \label{estim_eigen}
\hat{F} = \hatbar{V} \hat{\Gamma}_F \hatbar{V}^{\tp}, \quad \hat{G}_k = \hatbar{U}_k \hat{\Gamma}_{k} \hatbar{U}_k^{\tp} \quad (k=1,\ldots m).
\end{equation}
Element-wise, we have
\begin{equation*}
\hat{F}_{ij} = \sum_{\ell=1}^{\hat{d}_1} \gamma_{\ell}(\hat{F}) \hatbar{V}_{i\ell} \hatbar{V}_{j\ell} \quad (i=1,\ldots,n; \ j=1,\ldots n),
\end{equation*}
where $\hat{d}_1$ is the rank of $\hat{F}$, and $\gamma_{\ell}(\hat{F})$ denotes the $\ell$th eigenvalue of $\hat{F}$, ordered by magnitude. The elements of each $\hat{G}_k$ can be expressed similarly, with $\hat{d}_{2,k}$ denoting the rank of $\hat{G}_k$.
Then, fixing the estimated eigenvectors, the refitting step solves the convex problem
\begin{equation} \label{refit_objective}
\underset{\hat{\Gamma}_F,\hat{\Gamma}_k}{\min} \left\{ - \sum_{k=1}^m \sum_{i \leq j} \log Q \left( A_{k,ij}; \sum_{\ell=1}^{\hat{d}_1} \gamma_{\ell}(\hat{F}) \hatbar{V}_{i\ell} \hatbar{V}_{j\ell} + \sum_{\ell=1}^{\hat{d}_{2,k}} \gamma_{\ell}(\hat{G}_k) \hatbar{U}_{k,i\ell} \hatbar{U}_{k,j\ell} ,\phi \right) \right\}.
\end{equation}
When $Q$ is a one-parameter exponential family and the edge distribution is modeled through the corresponding canonical link function, see Example~\ref{model2},
solving \eqref{refit_objective} is exactly equivalent to fitting a generalized linear model with $n(n+1)m / 2$ responses and $\hat{d}_1 + \sum_{k=1}^m \hat{d}_{2,k}$ predictors.

With the solution to the refitting step problem \eqref{refit_objective}, we can construct the final estimates for the low-rank matrices based on these refitted eigenvalue estimates, along with the original estimated eigenvectors defined in \eqref{estim_eigen}.

\subsection{Choosing tuning parameters} \label{subsection_tuning}


A standard method for choosing tuning parameters is cross-validation, which requires some care on networks.   We take an approach motivated by the edge cross-validation for networks  \citep{li20network}, where a random subsample of node pairs is repeatedly removed, a low-rank matrix completion method is applied to the adjacency matrix to impute the missing pairs, and the original method is refit on the completed matrix. Tuning parameters are then selected to minimize a loss function evaluated on the held-out edges.

While the general edge cross-validation procedure \citep{li20network} contains an imputation step followed by a fitting step, MultiNeSS fitting approach can be applied directly to adjacency matrices with missing entries. Suppose we subsampled matrices $\{A_k\}_{k=1}^m$ by removing the values for a random sample of indices $(i,j,k)$, accounting for symmetry. Denote the set of remaining indices by $\Omega$. The new log-likelihood will resemble \eqref{log_likelihood}, but with the summation restricted to the triples in $\Omega$, and the same proximal gradient descent algorithm can be applied.


Similar to the approach taken by \cite{lock20bidimensional} for low-rank multiview data matrices, the tuning parameters can also be chosen adaptively using random matrix theory. In particular, in Example~\ref{model1} with known $\sigma$ constant across all layers, bounds on the singular values of $\sum_{k=1}^m E_k$ would suggest setting $\lambda = (2+\delta) \sigma \sqrt{nm}$ for a constant $\delta$.
\cite{gavish14optimal} introduce an estimator $\hat{\sigma}_{\textsc{MAD}}$ for $\sigma$ based on the median singular value and suggest setting $\delta = 0.309$, which is optimal for hard singular-value thresholding.
However, $\delta$ could also be selected using edge cross-validation.
Then, constant $\sigma$ across layers suggests the choice $\alpha_k = m^{-1/2} \ (k=1,\ldots, m)$.
This adaptive tuning scheme is used for the evaluation on synthetic networks in Section~\ref{section_simulations}.
While this approach is designed with Example~\ref{model1} in mind, it gives sensible results in Example~\ref{model2} with Bernoulli edges as well when the networks are sufficiently dense. For sparse networks with Bernoulli edges, we recommend setting $\lambda = C \sqrt{nm}$ and $\alpha_k = m^{-1/2}\ (k=1,...,m)$, where $C$ is a constant selected using edge cross-validation.

This adaptive tuning approach can also be used to account for layer-specific variances. Suppose $\hat{\sigma}^2_{\textsc{MAD}}(A_k)$ estimates the entry-wise variance for layer $k$.
Then rather than setting $\alpha_k$ the same for all layers, we set it based on the relative variance estimates for the different layers:
\begin{equation} 
  \alpha_k = m^{-1/2} \left\{ \frac{\hat{\sigma}_{\textsc{MAD}}(A_k)^2}{\sum_{k'=1}^m \hat{\sigma}_{\textsc{MAD}}(A_{k'})^2}\right\}^{1/2} \quad (k=1,\ldots ,m).
\end{equation}
As above, $\lambda$ is selected based on the singular values of $\sum_{k=1}^m E_k$,
\begin{equation} \label{adaptive_lambda}
  \lambda = (2+\delta) \sqrt{nm} \left\{ \sum_{k=1}^m \hat{\sigma}_{\textsc{MAD}}(A_{k})^2\right\}^{1/2},
\end{equation}
where again $\delta$ is a constant which is either chosen a priori or selected using edge cross-validation. This layer-specific adaptive tuning is used for the real data analysis in Section~\ref{section_realdata}.

The estimation algorithms described in this section for the models in Examples \ref{model1} and \ref{model2}, including options for refitting and parameter tuning, are implemented in an R package \texttt{multiness},
available at \texttt{github.com/peterwmacd/multiness}.

%% file: theory.tex
\section{Theoretical guarantees} \label{section_theory}

\subsection{Notation}

We denote the matrix $\ell_2$ operator norm by $\lVert M \rVert_2$. Let
\begin{equation*}
  [M]_d = \operatorname{argmin}_{M':\operatorname{rank}(M')\leq d} \lVert M - M' \rVert_F,
\end{equation*}
which is well-defined by the Eckart-Young Theorem as the truncation of the SVD of $M$ to the largest $d$ singular values.
For $d,p,q \geq 0$, let $\mathcal{O}_d$ denote the set of $d
\times d$ rotation (orthonormal) matrices, and $\mathcal{O}_{p,q}$
denote the set of $(p+q) \times (p+q)$ indefinite orthogonal
matrices. Let $\col(M)$ and $\row(M)$ denote the column and row spaces
of a matrix $M$, respectively.  For a symmetric matrix $M$, let
$\gamma_i(M)$ denote the $i$th eigenvalue of $M$, with eigenvalues
ordered from largest to smallest in absolute value. Throughout the
paper, any reference to ``leading'' or ``first'' eigenvalues of a
symmetric matrix refers to the largest in absolute value.

\subsection{Main results}

Throughout this section we assume the model described in
Example~\ref{model1}, where $Q(\cdot;\theta,\sigma)$ is the Gaussian
distribution with known variance $\sigma^2$.  To simplify notation, we assume that $d_2$ is constant in $k$, although the results generalize to the case where $d_{2,k}$ can depend on $k$, replacing $d_2$ in the assumptions by $\max_k d_{2,k}$.
We allow the dimensions $n$, $m$, $d_1$ and $d_2$ to grow, subject to the following restrictions:
\begin{assumption} \label{order_assumptions}
\begin{equation*}
d_2 2^{d_2} m^2 n^{1 - 2\tau} = o(1), \quad
d_1 m^{-1} = o(1), \quad
d_2 m^{-1} = o(1),
\end{equation*}
for some constant $\tau \in (1/2,1]$.
\end{assumption}
Assumption~\ref{order_assumptions} puts bounds on the total number of latent dimensions relative to the number of nodes $n$.
We study the estimator of the MultiNeSS model, defined as the limit of the proximal gradient update steps \eqref{update_theory}, starting from some initial value $\hat{F}^{(0)}$.
Let $\hat{F}$ and $\{ \hat{G}_k \}_{k=1}^m$ denote the limits of this proximal gradient descent algorithm as $t \rightarrow \infty$.

Simular to \eqref{estim_eigen}, let $G_k =  \bar{U}_k \Gamma_k \bar{U}_k \ (k=1,\ldots ,m)$
denote the eigen-decomposition of each $G_k$, and \\ $F = \bar{V} \Gamma_F \bar{V}^{\tp}$ denote the eigen-decomposition of $F$.
Suppose that they satisfy the following assumptions.

\begin{assumption} \label{eig_assumptions}


\begin{align} \label{eig_assump1}
b_1 n^{\tau} &\leq \lvert \gamma_{d_2}(G_k) \rvert \leq \lvert \gamma_1(G_k) \rvert = \lVert G_k \rVert_2 \leq B_1n^{\tau} \quad (k=1,\ldots ,m), \\
\label{eig_assumpF}
b_1 n^{\tau} &\leq \lvert \gamma_{d_1}(F) \rvert \leq \lvert \gamma_1(F) \rvert = \lVert F \rVert_2 \leq B_1n^{\tau}
\end{align}
for uniform constants $0 < b_1 \leq B_1$.

Further, assume
\begin{align} \label{eig_assump2}
\lVert \bar{V}^{\tp} \bar{U}_k \rVert_2 &= o(d_1^{-1/2} m^{1/2} n^{1/2 - \tau}) \quad (k=1,\ldots, m),
\end{align}
and
\begin{equation} \label{eig_assump_strong}
\lVert \bar{U}_{\mathcal{A}}^{\tp} \bar{U}_k \rVert_2 \leq B_2 \sigma \lvert \mathcal{A} \rvert^{1/2} n^{1/2 - \tau} \quad (k=1,\ldots, m)
\end{equation}
for some uniform constant $B_2 > 0$, where $\mathcal{A} \subseteq \{1,\ldots,m\} \setminus \{k\}$, and $\bar{U}_{\mathcal{A}}$ is an orthonormal basis for $\sum_{j \in \mathcal{A}} \col(G_j)$.

In particular,
\begin{equation} \label{eig_assump_weak}
\lVert \bar{U}^{\tp}_{k_1} \bar{U}_{k_2} \rVert _2 \leq B_2 \sigma n^{1/2 - \tau} \quad (k_1=1,\ldots,m;\ k_1=1,\ldots,m;\ k_1 \neq k_2).
\end{equation}
\end{assumption}

Although stated with fixed orthonormal bases,  \eqref{eig_assump2}, \eqref{eig_assump_strong} and \eqref{eig_assump_weak} are basis-free, and can be written in terms of the maximal cosine similarity between elements of the two column spaces. That is, if $S_1$ and $S_2$ are two subspaces of $\mathbb{R}^n$, then for any of their respective orthonormal bases $U_{S_1}$ and $U_{S_2}$,
\begin{equation*}
\lVert U_{S_1}^{\tp} U_{S_2} \rVert_2 = \sup_{x \in S_1,y \in S_2} \frac{ \lvert x^{\tp}y \rvert}{\lVert x \rVert_2 \lVert y \rVert_2}.
\end{equation*}
Comparing \eqref{eig_assump2} and \eqref{eig_assump_weak}, note that these conditions allow for slightly more similarity between the column spaces of $F$ and any one $G_k$ than between the column spaces of $G_k$ and $G_j$ for $k \neq j$.

Assumption~\ref{eig_assumptions} controls the signal strength through
the eigenvalues of $F$ and each $G_k$, and the separation between the
common and individual latent dimensions through bounds on the inner products of eigenvectors of $F$ and each $G_k$. As our framework treats the latent positions as deterministic, we make assumptions directly about these eigendecompositions rather than about the generative distribution of the latent positions.



With these assumptions we have the following consistency result. The proof is given in Appendix~\ref{appendix_thm1}.

\begin{theorem} \label{thm1}
Suppose $Q(\cdot;\theta,\sigma) = \mathcal{N}(\theta,\sigma^2)$, and Assumptions~\ref{order_assumptions}~and~\ref{eig_assumptions} hold.
Let $\lambda = 3 c_{\lambda} \sigma \sqrt{nm}$, and $\alpha_k = \left( c_{\lambda} \sqrt{m} \right)^{-1}\ (k=1,\ldots ,m)$, where $c_{\lambda}$ is a
universal constant.
Then with probability greater than $1 - (m+1)n e^{-C_0 n}$ for some universal constant $C_0 > 0$, the initializer
\begin{equation*}
    \hat{F}^{(0)} = \left[ \frac{1}{m} \sum_{k=1}^m A_k \right]_{d_1}
\end{equation*}
satisfies
\begin{equation} \label{init_rate}
    \lVert \hat{F}^{(0)} - F \rVert_F = o(n^{1/2}),
\end{equation}
and for $n$ sufficiently large, and all $k \in \{1,\ldots,m\}$, we have
\begin{equation} \label{thm1_bounds}
n^{-1} \lVert \hat{F} - F \rVert_F \leq C_1 \sigma d_1^{1/2}(nm)^{-1/2}, \quad
n^{-1} \lVert \hat{G}_k - G_k \rVert_F \leq C_2 \sigma d_2^{1/2} n^{-1/2} \quad (k=1,\ldots ,m)
\end{equation}
for positive constants $C_1$ and $C_2$ which do not depend on $n$, $m$, $d_1$, $d_2$, and $\sigma$.   Moreover, if all the eigenvalues of $F$ and each $G_k$ are non-negative, $\hat{F}$ and each $\hat{G}_k$ are positive semi-definite.
\end{theorem}

\begin{remark}
The initializer $\hat{F}^{(0)}$ uses the true value of $d_1$, which is generally unknown in practice. However, since the objective is convex, the estimators should not be sensitive to the initial value.
\end{remark}

\begin{remark} \label{optimal}
The conditions of Theorem~\ref{thm1} provide a regime under which our convex approach achieves the same rate as an oracle hard thresholding approach. In particular, if we estimated each $G_k$ with full knowledge of $F$, and $F$ with full knowledge of each $G_k$ by
\begin{equation*}
      \hat{F}^{(\text{oracle})} = \left[ \frac{1}{m} \sum_{k=1}^m (A_k - G_k) \right]_{d_1}, \quad \hat{G}_k^{(\text{oracle})} = \left[ A_k - F \right]_{d_2} \quad (k=1,\ldots ,m),
\end{equation*}
they would have the same Frobenius norm error rates as the estimators in Theorem~\ref{thm1}.
\end{remark}

\begin{remark} \label{usvt_errors}
In the proof of Theorem~\ref{thm1}, we will bound the operator norms of each error matrix $E_k$ using a concentration inequality for Gaussian random matrices \citep{bandeira16sharp}.  With a different operator norm concentration inequality \citep{chatterjee15matrix}, we can show that a similar result holds if the entries of $E_k$ are uniformly bounded instead of Gaussian. For instance, this would provide consistency for an RDPG-like binary edge model with $F + G_k \in [0,1]^{n \times n}$ for $k=1,\ldots,m$, and
\[
A_{k,ij} \sim \text{Bernoulli}(F_{ij} + G_{k,ij}) \quad (i=1,\ldots,n;\ j=1,\ldots, n;\ i < j; \ k=1,\ldots, n ).
\]
\end{remark}

While Assumption~\ref{order_assumptions} allows us to match the oracle error rate, it also places a strong requirement on the latent dimensions, especially the individual latent dimension $d_2$. Theorem~\ref{thm2} gives an alternative result under a weaker assumption on $d_2$, when it is allowed to grow polynomially in $n$. The proof is given in Appendix~\ref{appendix_thm2}.

\begin{assumption} \label{order_assumptions_weak}
    \begin{equation*}
        m^2 n^{1 - 2\tau} = o(1), \quad
        d_2 d_1 m^{-1} = o(1),
    \end{equation*}
    for some constant $\tau \in (1/2,1]$.
\end{assumption}

\begin{theorem} \label{thm2}
Suppose $Q(\cdot;\theta,\sigma) = \mathcal{N}(\theta,\sigma^2)$, and Assumptions~\ref{eig_assumptions}~and~\ref{order_assumptions_weak} hold.
Let $\lambda = 3 c_{\lambda} \sigma \sqrt{d_2 nm}$, and $\alpha_k = \left( c_{\lambda} \sqrt{d_2 m} \right)^{-1} \ (k=1,\ldots ,m)$, where $c_{\lambda}$ is a universal constant.  Then with probability greater than $1 - (m+1)n e^{-C_0 n}$ for some constant $C_0 >0$, the initializer
\begin{equation*}
    \hat{F}^{(0)} = \left[ \frac{1}{m} \sum_{k=1}^m A_k \right]_{d_1}
\end{equation*}
satisfies
\begin{equation*}
\lVert \hat{F}^{(0)} - F \rVert_F = o(n^{1/2}),
\end{equation*}
and for $n$ sufficiently large, we have
\begin{equation*}
n^{-1}\lVert \hat{F} - F \rVert_F \leq C_3 \sigma (d_1d_2)^{1/2} (nm)^{-1/2}, \quad
n^{-1} \lVert \hat{G}_k - G_k \rVert_F \leq C_4 \sigma d_2^{1/2}n^{-1/2} \quad (k=1,\ldots ,m)
\end{equation*}
for positive constants $C_3$ and $C_4$ which do not depend on $n$,
$m$, $d_1$, $d_2$, and $\sigma$. Moreover, if all the eigenvalues of $F$ and each $G_k$ are non-negative, $\hat{F}$ and each $\hat{G}_k$ are positive semi-definite.
\end{theorem}

Theorems~\ref{thm1} and \ref{thm2} provide bounds on the recovery of the $n \times n$ matrix-valued parameters $F$ and $G_k$, however in practice we are often interested in the latent position matrices $V$ and $U_k$ as well.
With an additional assumption on the eigenvalue gaps of $F$ and each $G_k$, the following Proposition~\ref{prop_ase_bound} establishes overall consistency for an adjacency spectral embedding-based estimate of the latent positions, after a suitable linear transformation.

Since in general, $\hat{F}$ and each $\hat{G}_k$ may have negative
eigenvalues, we define the adjacency spectral embedding (ASE) as in
\cite{rubin17statistical} based on the absolute values of the
eigenvalues.  
For instance, denoting the truncated eigendecomposition (up to rank
$d_1$) of $\hat{F}$ by $\hat{F} = \hatbar{V} \hat{\Gamma}_F
\hatbar{V}^{\tp}$, we define the $d_1$-dimensional ASE of $\hat{F}$  by $\hat{V} = \hatbar{V} \lvert \hat{\Gamma}_F \rvert^{1/2}$.

\begin{assumption} \label{eig_separation}
  \begin{equation}
    \operatorname{min}_{j \in \{2,\ldots,d_1\}} \left( \lvert \gamma_j(F) \rvert - \lvert \gamma_{j-1}(F) \rvert \right) \geq b_3 n^{\xi}
  \end{equation}
  for some $\xi \in (1/2,\tau]$ and positive constant $b_3$, and an analogous condition holds for the eigenvalues of each $G_k$ matrix with the same constant $\xi$.
\end{assumption}
This assumption on the eigenvalue gaps ensures that the ordering of
latent dimensions is preserved in the estimates of $F$ and of each $G_k$. We have the following consistency result for the latent matrices $V$ and of each $U_k$, up to rotation. The proof is given in Appendix~\ref{appendix_prop2}.

\begin{proposition} \label{prop_ase_bound}
  Suppose the assumptions of Theorem~\ref{thm1} and Assumption~\ref{eig_separation} hold. Then with probability greater than $1 - (m+1)n e^{-C_0 n}$ for some universal constant $C_0 > 0$, and for sufficiently large $n$, $\hat{F}$ and each $\hat{G}_k$ are low-rank matrices.
  Further, let $\hat{V}$ be the $n \times d_1$ dimensional ASE of $\hat{F}$, and $\hat{U}_k$ be the $n \times d_2$ dimensional ASE of $\hat{G}_k$ for each $k=1,\ldots,m$.
  Let $p_1$ and $q_1$ denote the number of assortative and disassortative common latent dimensions respectively, so that $F = VI_{p_1,q_1}V^{\tp}$. Define $p_2$ and $q_2$ similarly.
  Then we have 
  \begin{align}
      (d_1 n)^{-1/2} \operatorname{inf}_{W \in \mathcal{O}_{p_1,q_1}} \lVert \hat{V} - VW \rVert_F &\leq C_5 \sigma d_1^{1/2} m^{-1/2} n^{\tau/2 - \xi}, \label{ase_bound_F} \\
      (d_2 n)^{-1/2} \operatorname{inf}_{W \in \mathcal{O}_{p_2,q_2}} \lVert \hat{U}_k - U_kW \rVert_F &\leq C_6 \sigma d_2^{1/2} n^{\tau/2 - \xi} \quad (k=1,\ldots ,m) \label{ase_bound_G}
  \end{align}
  for some positive constants $C_5$ and $C_6$.
\end{proposition}

\begin{remark}
  Since we assume $\xi > 1/2 \geq \tau/2$, Proposition~\ref{prop_ase_bound} shows that under the asymptotic regime of Assumption~\ref{order_assumptions}, the average entry-wise error of the latent position matrices (after suitable linear transformation) goes to zero. As in Theorem~\ref{thm1}, the rate of convergence for the common structure exceeds that of the individual structure by a factor of $\sqrt{m}$.
\end{remark}

%% file: simulations.tex
\section{Evaluation on synthetic networks} \label{section_simulations}

\subsection{Baseline methods}

Throughout this section we compare the estimator for the MultiNeSS model
to other baseline methods on two types of synthetic networks:   with weighted edges generated according to the Gaussian model in Example~\ref{model1}, and with binary edges generated according to the logistic model in Example~\ref{model2}.
We compare to non-adaptive optimization approaches for the MultiNeSS model, and to other methods in the literature \citep{arroyo19inference, wang19common} for multiple networks which can capture the common or individual low-rank structure.

We also include two non-convex oracle approaches. For the Gaussian model, we apply a non-convex alternating rank truncation algorithm which assumes oracle knowledge of the true ranks $d_1$ and $d_2$.
The alternating updates for $t \geq 1$ are given by
\begin{equation*}
\hat{F}^{(t)} = \left[ \frac{1}{m} \sum_{k=1}^m \left( A_k - \hat{G}_k^{(t-1)} \right) \right]_{d_1}, \quad
\hat{G}_k^{(t)} = \left[ A_k - \hat{F}^{(t)} \right]_{d_2} \quad (k=1,\ldots,m),
\end{equation*}
and initialized with $\hat{G}_k^{(0)} = 0\ (k=1,\ldots,m)$.
These update steps are applied until convergence, or until a pre-specified maximum iteration number $t_{\max}$ is reached.

For the logistic model we compare our convex approach with a non-convex gradient descent algorithm, similar to \cite{ma20universal}, which also assumes known $d_1$ and $d_2$.  This approach directly updates the entries of the latent position matrices $V$ and each $U_k$ by performing gradient descent on the negative log-likelihood function.


The recently proposed COSIE method \citep{arroyo19inference} fits a low-rank model to multiple binary undirected networks on a common node set.
COSIE provides estimates of the expected adjacency matrices for each layer, but does not decompose the estimate into common and individual parts, so we can only compare the accuracy of overall expectation.
While COSIE is designed for the RDPG model, it can also be applied unchanged to the Gaussian model. We apply an oracle version of COSIE assuming knowledge of the true $d_1$ and $d_2$.
For a fair comparison to our method,  we first identify the $d_1+d_2$ leading eigenvectors for each layer, then use COSIE to fit a common invariant subspace of dimension $d_1 + md_2$, the total number of latent dimensions in the MultiNeSS model.

The second baseline comparison is to the M-GRAF algorithm proposed by \cite{wang19common}, for a similar logistic link model for multilayer networks with common and individual parts.
The M-GRAF model does not assume any structure, low-rank or otherwise, for entries of the common matrix $F$ and does not employ regularization, and is thus better suited to the regime with small $n$ and large $m$.
We apply an oracle version of M-GRAF which assumes knowledge of the true individual rank $d_2$. Since M-GRAF does not assume a common low-rank structure, it does not need a value for $d_1$.

\subsection{Gaussian model results} \label{section_gaussiansims}

We consider instances of the Gaussian model with no self-loops, the usual inner product similarity, $d_1=d_2=2$, and $\sigma=1$, where we vary $n \in \{200,300,400,500,600\}$ with fixed $m=8$, and vary $m \in \{4,8,12,15,20,30\}$ with fixed $n=400$.
In each setting we generate 100 independent realizations from the model.
The entries of the common and individual latent position matrices are generated as independent standard normals, so while they are not strictly orthogonal, their expected correlation is 0.
Under the Gaussian model, we have four methods to compare: the MultiNeSS estimator with and without the refitting step, denoted MultiNeSS and MultiNeSS+, respectively; the alternating rank truncation approach (Non-convex); and COSIE.

We evaluate the methods on how well they do on recovering the common structure, the individual structure, and the overall expectation of the adjacency matrix, using relative Frobenius norm errors with  $\lVert \cdot \rVert_{\tilde{F}}$ denoting the Frobenius norm which ignores diagonal entries:
\begin{align}
  \label{error_fg}
  \mathrm{Err}_{F} 
  = \frac{\lVert \hat{F} - F \rVert_{\tilde{F}}}{\lVert F \rVert_{\tilde{F}}}, &\quad
  \mathrm{Err}_{G}
  = \frac{1}{m} \sum_{k=1}^m \frac{ \lVert \hat{G}_k - G_k \rVert_{\tilde{F}}}{\lVert G_k \rVert_{\tilde{F}}} , \\
  \mathrm{Err}_{P} 
  =  \frac{1}{m} &\sum_{k=1}^m \frac{ \lVert \hat{F} + \hat{G}_k - F - G_k \rVert_{\tilde{F}}}{\lVert F + G_k \rVert_{\tilde{F}}}. \nonumber
\end{align}


\begin{figure}
    \centering
    \includegraphics[width=.8\textwidth]{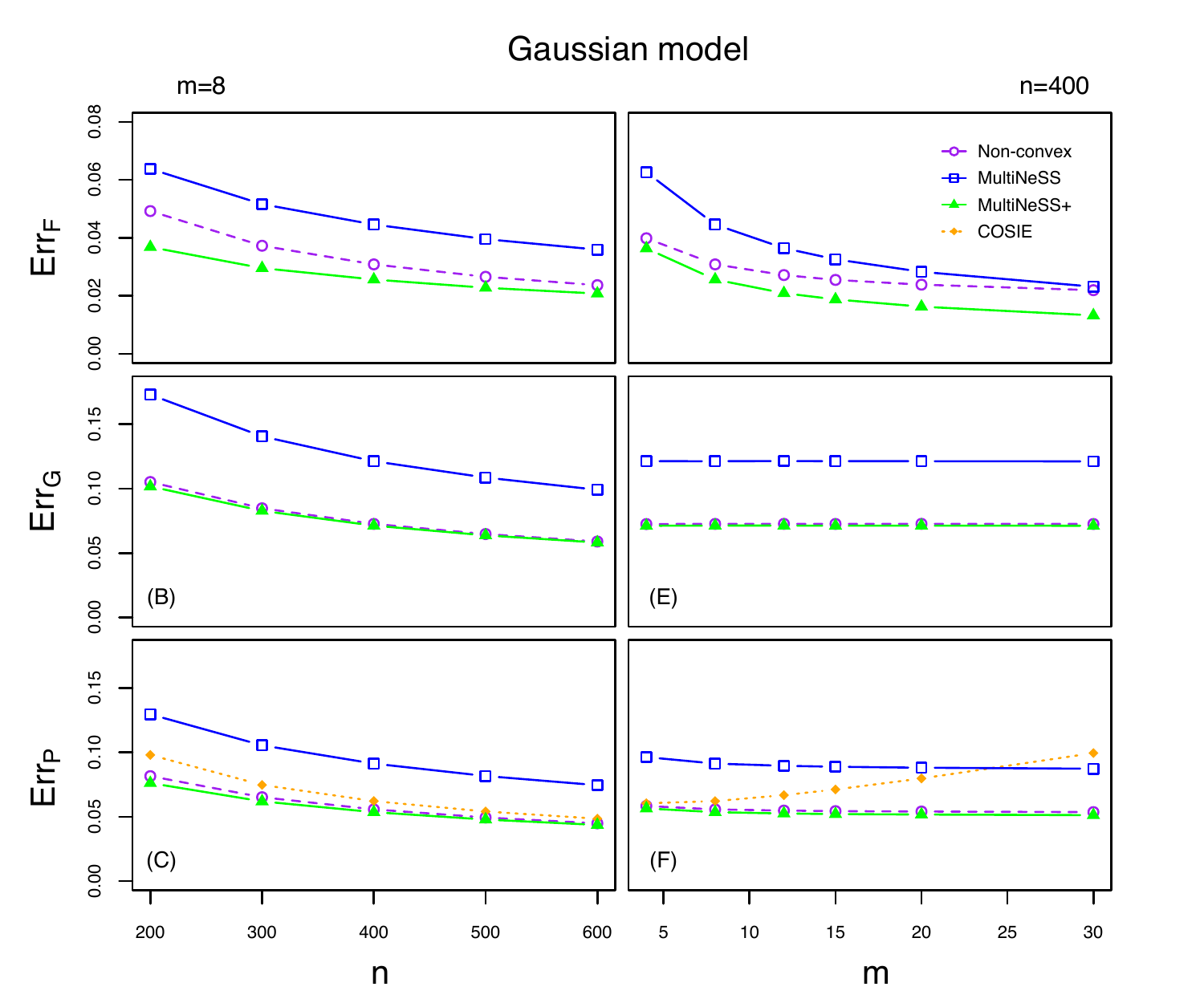}
    \caption{Frobenius norm errors for the common structure (top row),  individual structure (middle row) and overall expected value (bottom row) under the Gaussian model.}
    \label{gaussian_plot}
\end{figure}

The results are shown in  Figure~\ref{gaussian_plot}.  Panels (A), (B), and (C) on the left show the errors as a function of the number of nodes $n$, with the fixed number of layers $m = 8$.   Panels (D), (E), and (F) on the right show the errors as the number of layers $m$ increases, with $n = 400$ fixed.

There are several general conclusions to draw here. MultiNeSS without the refitting step does not outperform the non-convex oracle, but MultiNeSS+ is uniformly the best method in all cases, though the non-convex oracle performs very similarly on estimating the individual layers $G_k$.  One possible explanation for improvement over the non-convex oracle is that the convex optimization approach ignores the diagonal elements of the adjacency matrices, which do not reflect the true low-rank structure.

All methods perform better as the number of nodes $n$ grows, as we would expect. The number of layers $m$ growing has no effect on errors in estimating the individual components for MultiNeSS, since each one is estimated separately, but it helps us estimate $F$ better by pooling shared information across more layers and therefore also improves the overall estimation of $P$.
The rate of decrease in error in $F$ seems to match well the rate of $m^{-1/2}$ predicted by the theory.
COSIE, on the other hand, benefits from growing $n$ but suffers when $m$ grows, with the overall error in $P$ going up with $m$.   We conjecture that this happens because  COSIE must first estimate a subspace of dimension $d_1 + md_2$, which leads to high variability as $m$ grows.

Comparing panel (C) to panels (A) and (B), and panel (F) to panels (D) and (E), we see that the estimation error for $P_k$ is on average less than the estimation error for $G_k$, implying that the error in $P_k$ does not decompose additively into error for $G_k$ and error for $F$.
Even when the expected correlation in the latent position matrices is zero,
it is challenging to correctly distinguish common structure from individual structures.

\subsection{Logistic model results}

We also consider instances of the logistic model with no self-loops, the same inner product similarity, and $d_1=d_2=2$, where we vary $n \in \{200,300,400,500,600\}$ with fixed $m=8$, and vary $m \in \{4,8,12,15,20,30\}$ with fixed $n=400$.
In each setting we generate 100 independent realizations of the model.
The entries of the common and individual latent position matrices are generated as independent standard normals.
We compare the MultiNeSS estimator with and without the refitting step (again denoted by MultiNeSS and MultiNeSS+) to the non-convex approach, COSIE, and M-GRAF.
Note that COSIE does not use the correct model for this data since it assumes a random dot product graph model without a logistic link.
We evaluate the recovery of the common and individual structures using the same relative Frobenius norm errors \eqref{error_fg}.
To evaluate the overall recovery of the expected value for each layer, we use the relative Frobenius norm error after element-wise application of the inverse logistic link function.
That is, we redefine
\begin{equation*}
\mathrm{Err}_{P} 
=  \frac{1}{m} \sum_{k=1}^m \frac{ \lVert g(\hat{F} + \hat{G}_k) - g(F + G_k) \rVert_{\tilde{F}}}{\lVert g(F + G_k) \rVert_{\tilde{F}}},
\end{equation*}
where $g$ is defined in \eqref{expit}.


\begin{figure}
    \centering
    \includegraphics[width=.8\textwidth]{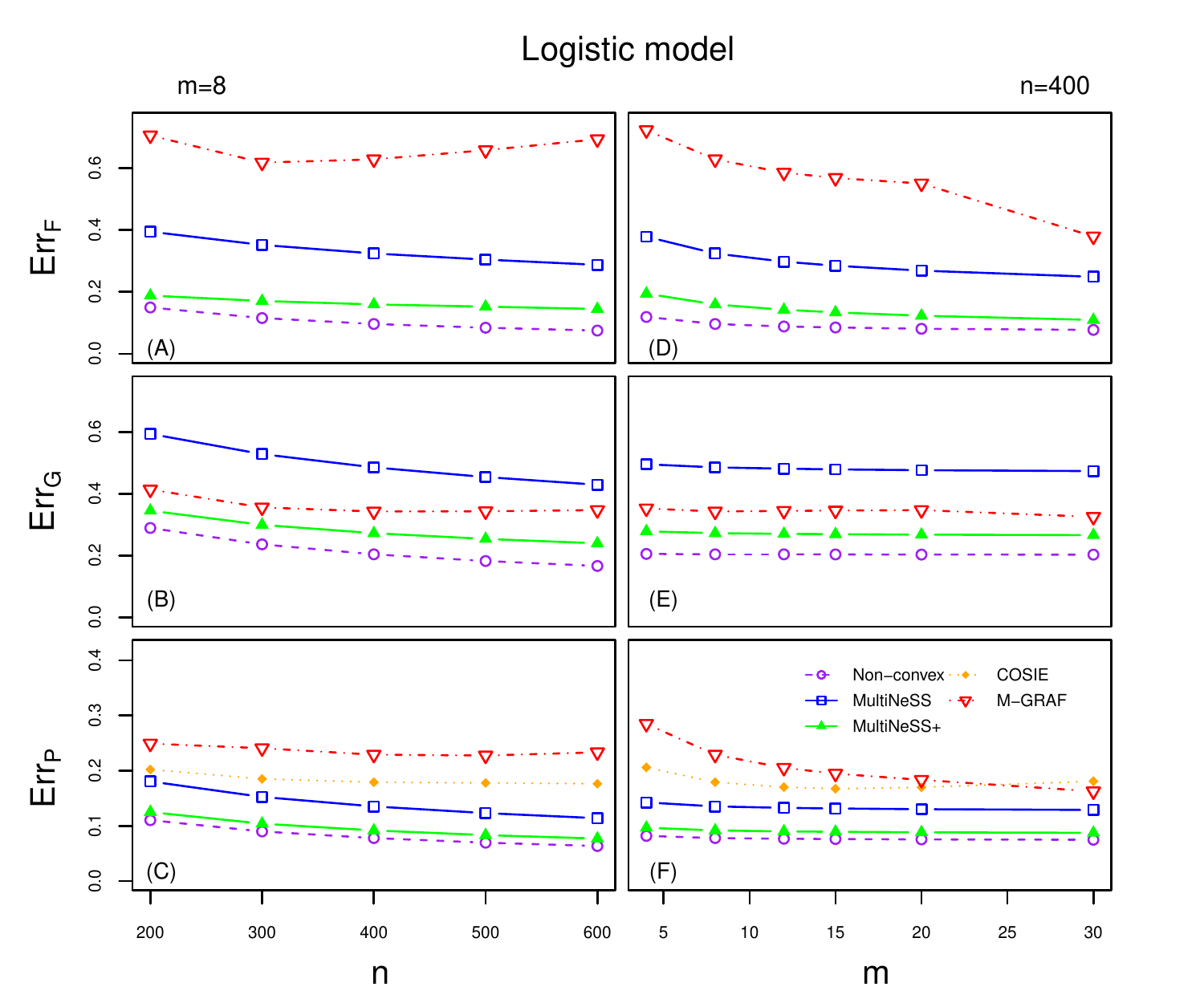}
    \caption{Frobenius norm errors for the common structure (top row),  individual structure (middle row), and the overall expected value after the logistic transformation(bottom row), under the logistic model.}
    \label{logistic_plot}
\end{figure}

The results are shown in  Figure~\ref{logistic_plot}.  Panels (A), (B), and (C) on the left show the errors as a function of the number of nodes $n$, with the fixed number of layers $m = 8$.   Panels (D), (E), and (F) on the right show the errors as the number of layers $m$ increases, with $n = 400$ fixed.

Many of the general conclusions here are the same as for the Gaussian model.
M-GRAF does not perform better as $n$ increases, and performs much worse for small values of $m$, since it does not regularize the common matrix $F$ in any way.
In contrast to the Gaussian model, here the non-convex approach slightly outperforms MultiNeSS+.
The difference between the non-convex and MultiNeSS+ errors is driven by large magnitude entries in $F$ and $G_k$ which have a substantial effect on the log-odds scale, but little effect on the expectation of the adjacency matrix.
Hence, the difference between these two methods is attenuated in panels (C) and (F) after applying the inverse logistic link function.


For the binary networks generated from the logistic MultiNeSS model, we also compare performance over a range of network edge densities by subtracting a density controlling parameter $\beta \geq 0$ from the log-odds of each edge.
As above, we generate $V$ and $\{U_k\}_{k=1}^m$ as $n \times 2$ matrices of independent standard normals, resulting in $P_k = g(VV^{\tp} + U_kU_k^{\tp} - \beta \bm{1}_n\bm{1}_n^{\tp})$.
This is equivalent to generating networks from a logistic MultiNeSS model with generalized inner product similarity, augmenting the common latent position matrix with an extra dissasortative latent dimension with coordinates $\sqrt{\beta} \bm{1}_n$.

We consider instances of this logistic MultiNeSS model with no self-loops, $n=400$, $m=8$, and $\beta \in \{0,1,2,3,4,5,6\}$. MultiNeSS networks generated with these choices of $\beta$ have edge densities of approximately $0.5,0.34,0.21,0.12,0.06,0.035,$ and $0.015$ respectively.
We compare the MultiNeSS estimator with and without the refitting step to the non-convex approach, COSIE, and M-GRAF.
In order to easily implement the non-convex oracle approach, we fit it with full knowledge of $d_1,d_2$, and $\beta$.  MultiNeSS without refitting is tuned adaptively with a fixed constant, as in the previous dense network simulations. MultiNeSS+ is tuned with edge cross-validation.
The error $\mathrm{Err}_F$ for the recovery of the common structure is normalized by $\lVert VV^{\tp} \rVert^2_{\tilde{F}}$, ignoring the effect of $\beta$ on the common structure; $\mathrm{Err}_P$ is calculated as above, including $\beta$ in the normalizer.

\begin{figure}
    \centering
    \includegraphics[width=.8\textwidth]{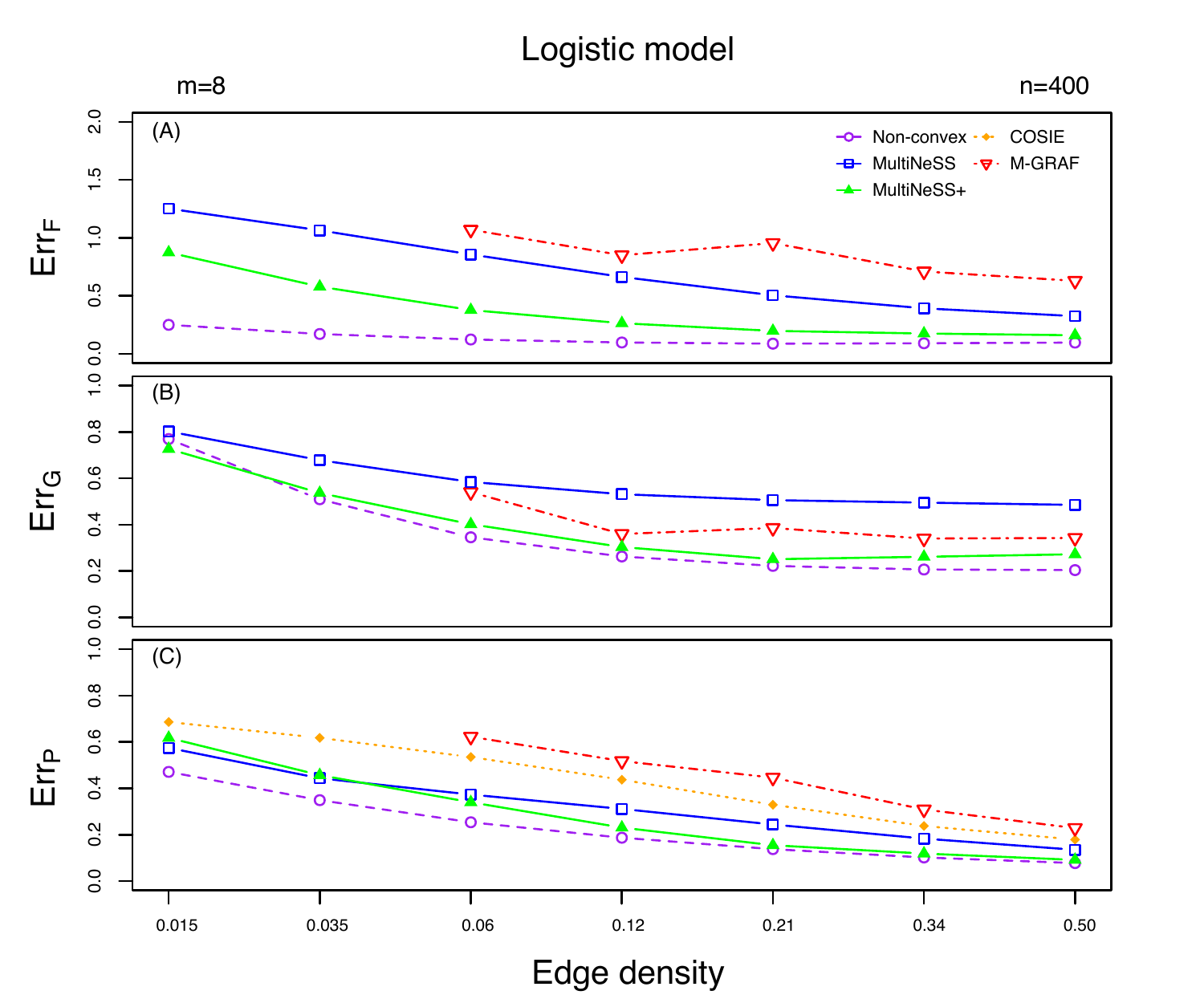}
    \caption{Frobenius norm errors for the common structure (top panel),  individual structure (middle panel), and the overall expected value
 (bottom panel), under the logistic model with varying edge density.}
    \label{logistic_density_plot}
\end{figure}

The results are shown in  Figure~\ref{logistic_density_plot}. For highly sparse networks with edge densities approximately 3.5\% and 1.5\%, M-GRAF does not converge consistently, so its results in these settings are omitted.

For edge densities over approximately 5\%, the relative performances of the methods are similar to those seen for dense networks.
As edge density decreases, the non-convex oracle unsurprisingly performs much better than MultiNeSS in panel (A), as it does not have to fit the density controlling parameter $\beta$.
For highly sparse networksw with edge densities under 5\%, we see that MultiNeSS+ outperforms MultiNeSS without refitting in panels (A) and (B), but has slightly worse error in panel (C). While MultiNeSS+ better controls the ranks of $F$ and $G_k$ and more accurately recovers the latent coordinates, MultiNeSS without refitting performs best with a much smaller choice of $\lambda$ and can more accurately recover the expected adjacency matrix despite greatly overestimating the number of latent dimensions.
Finally, we see that in the sparsest regime in panel (B), MultiNeSS+ outperforms the non-convex oracle.
In this case, MultiNeSS+ is able to adaptively ignore some weak latent dimensions, while the oracle non-convex approach is forced to fit two individual latent dimensions per layer, even when the signal is too weak to reliably estimate its coordinates.

%% file: real_data.tex
\section{An agricultural trade network analysis} \label{section_realdata}


As an illustration of insights one can gain from fitting a MultiNeSS model, we analyze a data set of food and agriculture trade relationships between countries, collected in 2010.
Each node corresponds to a country, and each layer to a different agricultural product. The undirected edges are weighted by the bilateral traded quantity of the commodity. This data set has previously been analyzed by \cite{de15structural}, who looked at structural similarities between layers.

As a pre-processing step, we remove low density layers and nodes. The original dataset contains 214 countries and 364 products.   We kept layers with at least 10\% non-zero edges, and included nodes with a mean of at least 5 non-zero edges across these layers. The result is an undirected multiplex network with no self loops, with $n=145$ nodes and $m=13$ layers corresponding to agricultural products with high trade volume. Following common practice for this type of data, we work with log trade volumes as edge weights, which also makes the assumption of Gaussian edge weights with constant variance within each layer more realistic.


We fit a Gaussian model using the MultiNeSS algorithm with refitting. The tuning parameters are selected using the layer-specific adaptive tuning approach described in Section~\ref{subsection_tuning}. The constant in \eqref{adaptive_lambda} is set to $\delta = 1/2$ using edge cross-validation.

\begin{figure}
    \centering
    \includegraphics[width=.8\textwidth]{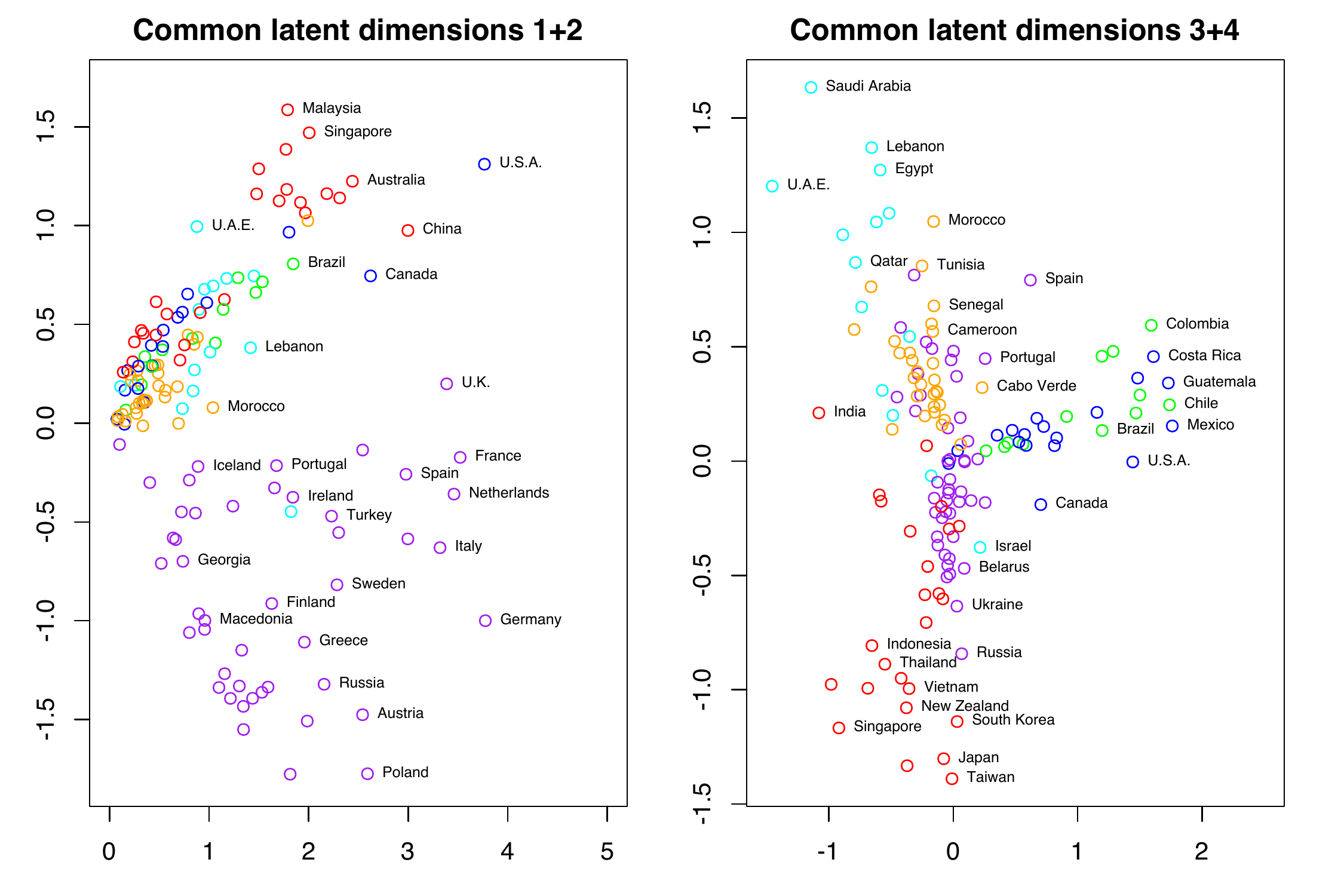}
    \caption{Scatter plots of first four MultiNeSS common latent dimensions of the food trade data. Left panel: dimensions 1 and 2, right panel: dimensions 3 and 4. Dimensions are colored black for assortative, and red for disassortative.  Points are colored by geographical region. Orange: Africa, Red: Asia-Pacific, Purple: Europe, Cyan: Mid East, Blue: North America, Green: South America.}
    \label{scatter_common}
\end{figure}

\begin{figure}
    \centering
    \includegraphics[width=.8\textwidth]{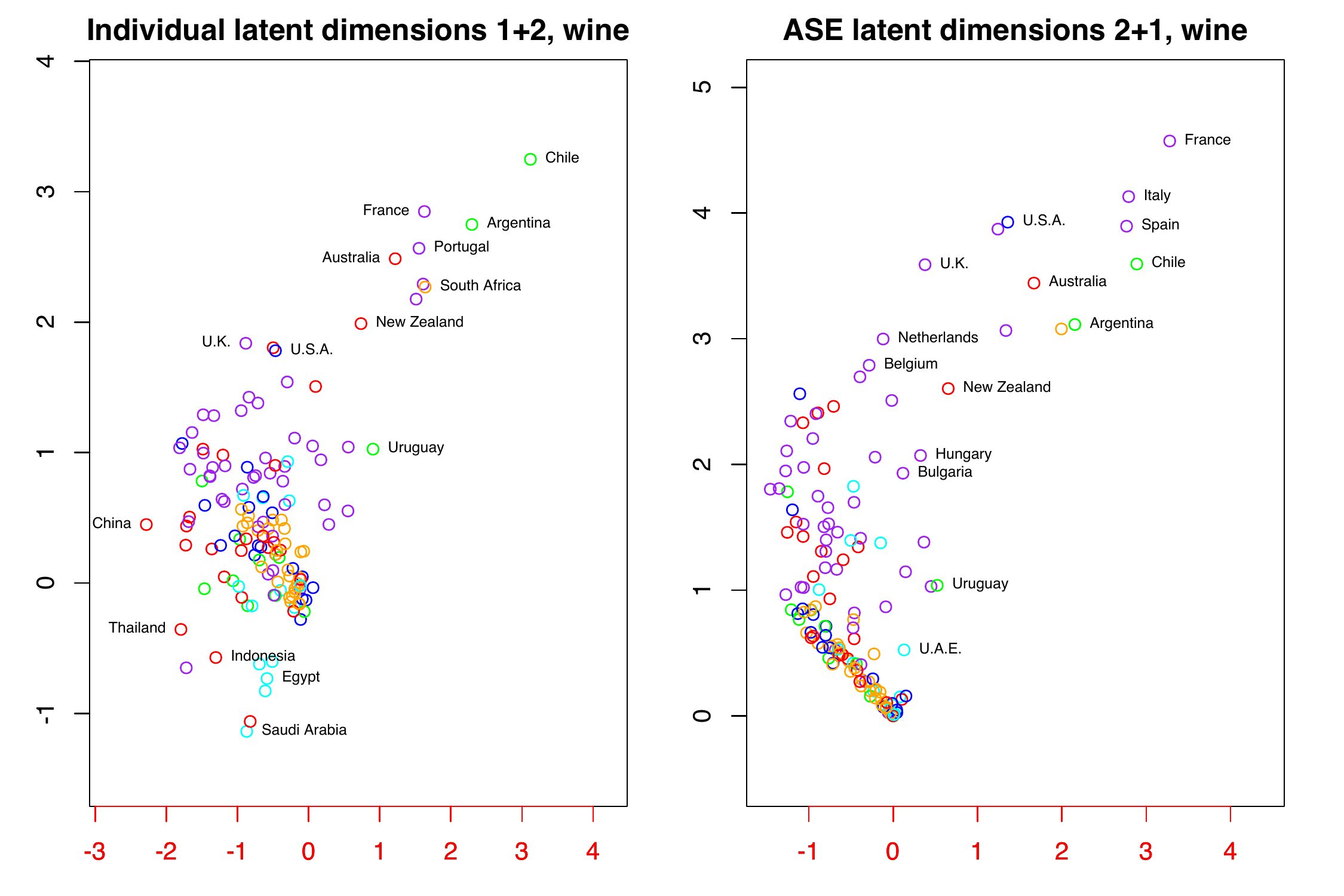}
    \caption{Scatter plots of first two individual latent dimensions, wine layer. Left panel: dimensions 1 and 2 by MultiNeSS, right panel: dimensions 1 and 2 by adjacency spectral embedding.  Dimensions are colored black for assortative, and red for disassortative.  Points are colored by geographical region. Orange: Africa, Red: Asia-Pacific, Purple: Europe, Cyan: Middle East, Blue: North America, Green: South America.}
    \label{scatter_wine}
\end{figure}

\begin{figure}
    \centering
    \includegraphics[width=.8\textwidth]{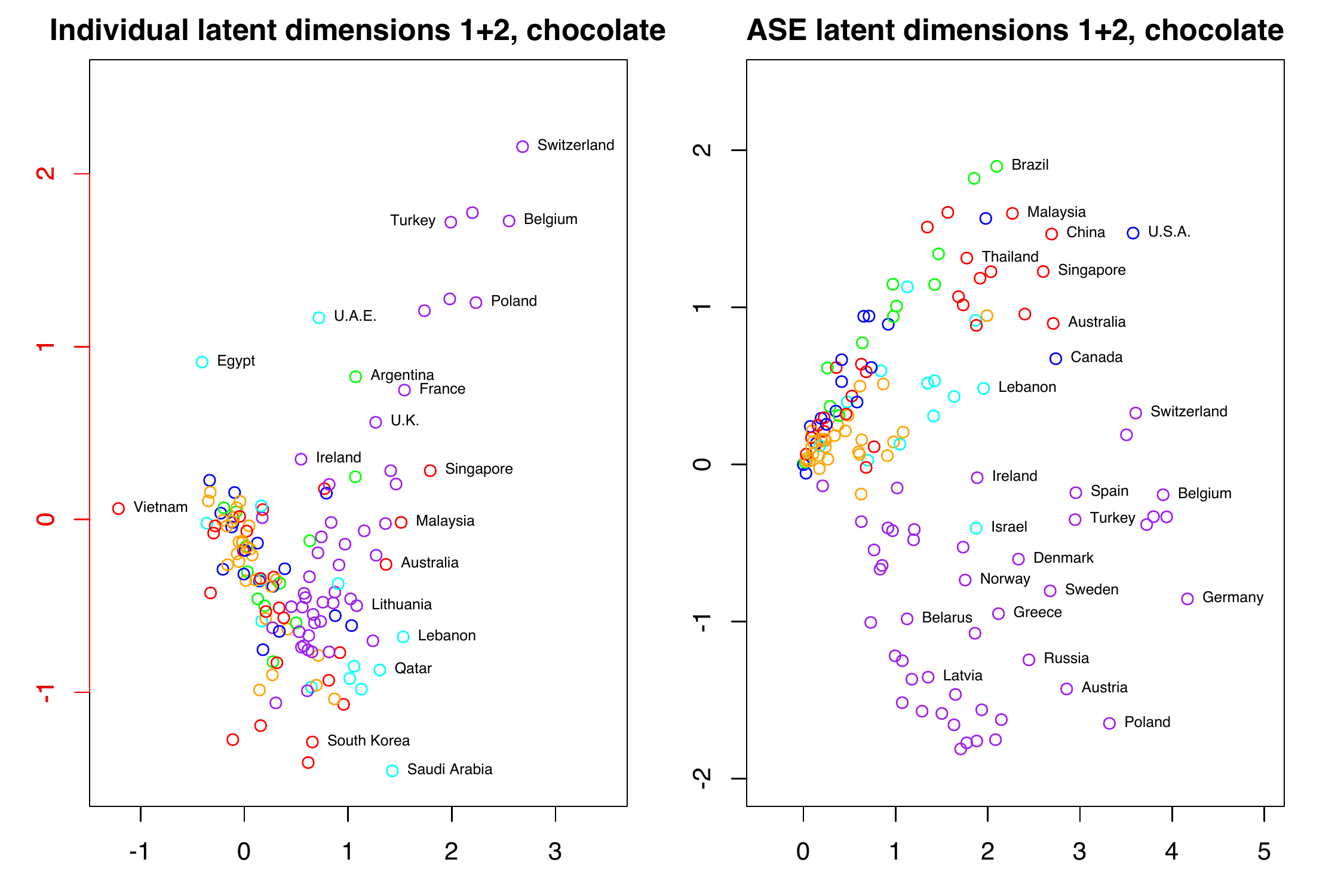}
    \caption{Scatter plots of first two individual latent dimensions, chocolate layer. Left panel: dimensions 1 and 2 by MultiNeSS, right panel: dimensions 1 and 2 by adjacency spectral embedding.   Dimensions are colored black for assortative, and red for disassortative.  Points are colored by geographical region. Orange: Africa, Red: Asia-Pacific, Purple: Europe, Cyan: Mid East, Blue: North America, Green: South America. }
    \label{scatter_chocolate}
\end{figure}

We show the results for the first four common latent dimensions in Figure~\ref{scatter_common}, and the first two individual dimensions for two example layers, wine and chocolate in Figure~\ref{scatter_wine} and Figure~\ref{scatter_chocolate} respectively.  



The estimated common matrix $\hat{F}$ has rank 39, with 25 assortative dimensions and 14 disassortative dimensions. Figure~\ref{scatter_common} shows the scatter plots of the points projected onto the leading four latent dimensions, first and second on the left, third and fourth on the right, which are all assortative.  The first four singular values account for approximately 47\% of the sum of the singular values of $\hat{F}$. The scatter plots suggest that the first latent dimension corresponds roughly to the total volume of trade, and the subsequent ones correspond to regional trade relationships. In particular, the second dimension primarily separates Europe from Asia, the third separates the Americas from the rest of the world, and the fourth separates the Middle East and Africa from Asia and the Pacific.


For the individual component of the wine trade layer, shown in Figure~\ref{scatter_wine}, we estimate $\rank(\hat{G}_{{\rm wine}}) = 18$, with 9 assortative and 9 disassortative latent dimensions. We plot the coordinates the first two latent dimensions, which account for about 37\% of the sum of the singular values of $\hat{G}_{{\rm wine}}$. The second latent dimension corresponds roughly to the total volume of wine production after correcting for the common structure, with countries like France, Spain, Chile, and New Zealand having very high scores, and majority Muslim nations like Saudi Arabia and Indonesia having very low scores.
The first latent dimension is dissasortative, and gives large positive coordinates to the major wine exporters who do not trade wine amongst themselves, separating them from major wine importers such as China.

For comparison, we also plot the countries projected onto the first two latent dimensions constructed by ASE applied to just the wine layer of the trade network.   We swap the order of the ASE dimensions to ease visual  comparison to the MultiNeSS embedding. While   the ASE dimensions have similar interpretations to MultiNeSS, and provide the same general conclusions about high volume wine producers, the interpretation of the lower-left part of the scatter plot is much more clear in the MultiNeSS individual embedding.



For the chocolate trading network, we estimate the individual component rank as $\rank(\hat{G}_{{\rm chocolate}}) = 7$, with 5 assortative and 2 disassortative latent dimensions.  Projections on to the first two individual latent dimensions, which account for about 48\% of the sum of the singular values of $\hat{G}_{{\rm chocolate}}$, are shown in Figure~\ref{scatter_chocolate}.  Overall, the pattern is similar to the left panel of Figure~\ref{scatter_wine}, with the two axes swapped. The first latent dimension identifies chocolate producing nations like Switzerland and Belgium with the highest scores, and countries like Vietnam, which has a very low per capita chocolate consumption, with the lowest score.
The second latent dimension is disassortative, and gives large positive coordinates to major chocolate exporters which do not trade chocolate with each other. Egypt and UAE also have outlying coordinates as according to this data, they primarily trade chocolate with other Middle Eastern nations rather than importing directly from Europe. The ASE in the right panel of Figure~\ref{scatter_chocolate} looks very similar to the left panel of Figure~\ref{scatter_common}. Since this embedding does not account for the common structure, it primarily captures patterns common to all products, rather than the chocolate-specific patterns revealed by the MultiNeSS embedding.

%% file: conclusion.tex
\section{Discussion}

\ifbka
\else
The central contribution of this work is MultiNeSS, a latent space model for multiplex networks with shared structure which allows for learning  both common and individual structure in the layers.    The model can be fitted with a convex optimization algorithm, and an additional fast de-biasing step can improve its fit.   The algorithm is data-driven and can adapt to different levels of noise at different edges.    We allow for general edge weight distributions and general similarity functions between latent positions as the center parameter of the distribution.    For the case of similarity measured by generalized inner product, we prove the identifiability of the model under a mild linear independence condition that does not require orthogonality of the common and individual latent dimension, and for the Gaussian model of edge weights, we establish consistency of our estimators.    We expect this can be extended to other well-behaved edge weight distributions.
We demonstrate the method's effectiveness over existing methods on simulated multiplex networks and on a food trading network, where it produces interpretable insights distinct from what one can get from separate analysis.
\fi



There are several directions in which we plan to take this work forward.
The models we developed so far allow for only two kinds of latent dimensions: those which are individual to one layer, and those which are common to all layers.   Extending this to more structured models, where latent dimensions can be shared by some but not all layers, would allow for a larger range of applications.
For example, we could impose a group structure on the layers, allowing for group effects and enabling an analogue to analysis of variance on networks.    An example application where this would be very useful is neuroimaging, where brain connectivity networks of a treatment group and a control group of patients could be analyzed jointly and the treatment effect estimated more accurately.      These groups could also be learned from data, in a natural extension of this setup to clustering.

Another possible extension  is to dynamic networks, where each layer represents a network snapshot at a discrete time point.  In this setting, unlike in ours, the ordering of the layers matters.   Latent dimensions could be modeled as constant over time or constant over a contiguous time window, with obvious applications to prediction and change-point analysis.
Finally, a highly interpretable latent structure could be obtained if we imposed a tree structure on the latent dimensions, with shared latent dimensions between nodes determined by their last common ancestor on the tree.

While this work focuses on undirected networks, we also recognize the importance of extending this model to directed networks. 
  In this case each node would have both incoming and outgoing coordinates for each latent dimension. For instance, we could model the common structure as $V_{{\rm out}}V_{{\rm in}}^{\tp}$ for $n \times d_1$ matrices $V_{{\rm in}}$ and $V_{{\rm out}}$. This directed model further complicates identifiability and interpretation. There is now a scale unidentifiability for each latent dimension, which means we cannot distinguish between the contributions of incoming and outgoing node behavior.

Finally, recent work has demonstrated that linear embeddings, which assume a low-rank structure on expected adjacency matrices, may be too restrictive to model complex real data \citep{rubin20manifold}. In our data application, we find our latent embedding of worldwide agricultural trade to have relatively high dimension compared to the number of nodes. There could be potential to further reduce the latent dimension by applying additional manifold dimension reduction to the common and individual embeddings.


%% file: appendix_optimization.tex

\ifbka
\else
  \section{Details of proximal gradient descent} \label{appendix_pgd}
\fi

Recall that we split the optimization variables into $m+1$ blocks of $n(n-1)/2$ variables: one block containing the entries of $F$; and one block each for the entries of each $G_k$. For each block indexed, starting from zero, by $j=0,\ldots,m$, define $h_{j,1}$, the smooth part of \eqref{ls_objective} written as a function of block $j$, and $h_{j,2}$, the part of the penalty which depends on block $j$:
\begin{align*}
h_{0,1}(F) + h_{0,2}(F) & = -\frac{1}{2}\sum_{m=1}^M \sum_{i \neq j} \log Q(A_{k,ij};F_{ij} + G_{k,ij}) + \lambda ||F||_*, \\
h_{k,1}(G) + h_{k,2}(G) & = -\frac{1}{2}\sum_{i \neq j} \log Q(A_{k,ij};F_{ij} + G_{k,ij})  + \lambda \alpha_k ||G||_*   \  (k=1,\ldots,m).
\end{align*}
Then for each $h_{k,1}$ is convex and differentiable, and $h_{k,2}$ is convex and, although non-differentiable, has a closed form proximal mapping for step size $\eta$ \citep{fithian18flexible}. In particular, the nuclear norm scaled by $\lambda \geq 0$ has the proximal mapping
\begin{equation*}
\underset{M'}{\text{argmin}} \frac{1}{2\eta} ||M - M'||_F^2 + \lambda||M||_* =S_{\eta\lambda}(M),
\end{equation*}
where $S_{T}(\cdot)$ is the soft singular value thresholding operator with threshold $T \geq 0$.
Differentiating $h_{k,1}$ gives
\begin{align*}
\frac{\partial h_{0,1}}{\partial F_{ij}} &= -\sum_{k=1}^m (\log Q)'(A_{k,ij};F_{ij}+G_{k,ij}) \\
\frac{\partial h_{k,1}}{\partial G_{k,ij}} &= -(\log Q)'(A_{k,ij};F_{ij}+G_{k,ij}) \quad (k=1,\ldots,m)
\end{align*}
if $i \neq j$, and
\begin{equation*}
    \frac{\partial h_{0,1}}{\partial F_{ii}} = \frac{\partial h_{k,1}}{\partial G_{k,ii}} = 0
\end{equation*}
otherwise. We select the relative step sizes based on the Lipschitz constants of each block objective. If $\log Q$ is Lipschitz in $\mu$ with constant $L$, then $h_{0,1}$ has Lipschitz constant $L_0= mL$, while each $h_{k,1}$ has Lipschitz constant $L_k \equiv L$. Thus proximal gradient descent with scaled step size $L\eta/L_k$ gives the following update steps: for iteration number $t \geq 1$,
\begin{align*}
\hat{F}^{(t)} &= S_{\eta\lambda/m} \left( \hat{F}^{(t-1)} - \frac{\eta}{m}\frac{\partial h_{0,1}(\hat{F}^{(t-1)})}{\partial F} \right), \\
\hat{G}_k^{(t)} &= S_{\eta\lambda\alpha_k}\left(\hat{G}_k^{(t-1)} - \eta \frac{\partial h_{k,1}(\hat{G}_k^{(t-1)})}{\partial G_k}\right) \quad (k=1,\ldots,m).
\end{align*}

%% file: appendix_proofs.tex

\ifbka
\else
  \section{Technical Proofs}
\fi

\setcounter{subsection}{0}
\subsection{Proof of Proposition~\ref{identifiability}} \label{appendix_prop1}


We begin with an auxilliary lemma that establishes identifiability of single-layer latent positions up to indefinite orthogonal transformation under the generalized inner product similarity. We then use a linear independence argument to establish Proposition~\ref{identifiability}.

\begin{lemma} \label{lemma_indef_rotation}

Suppose $X$ and $Z$ are $n \times d$ matrices of full column rank, such that for some non-negative integers $p,q$ satisfying $p+q=d$,
\begin{equation} \label{xz_outer}
  X I_{p,q} X^{\tp} = Z I_{p,q} Z^{\tp}.
\end{equation}
Then $X = ZW$ for an indefinite orthogonal rotation $W \in \mathcal{O}_{p,q}$.
\end{lemma}

\ifbka
  \begin{proof}[of Lemma~\ref{lemma_indef_rotation}]
\else
  \begin{proof}[Proof of Lemma~\ref{lemma_indef_rotation}]
\fi
  Since $X$ has full column rank $X^{\tp}X$ is invertible. Thus, by \eqref{xz_outer},
  \begin{align*}
    X &= X I_{p,q} X^{\tp} X (X^{\tp} X)^{-1} I_{p,q} \\
     &= Z I_{p,q} Z^{\tp} X (X^{\tp}X)^{-1} I_{p,q}.
  \end{align*}
  Denote $W = I_{p,q} Z^{\tp} X (X^{\tp}X)^{-1} I_{p,q}$. The proof is complete if we can establish that $W \in \mathcal{O}_{p,q}$. By \cite{rubin17statistical}, it suffices to show that $W^{\tp}I_{p,q}W = I_{p,q}$. Again using \eqref{xz_outer},
  \begin{align*}
    W^{\tp}I_{p,q}W &= I_{p,q} (X^{\tp}X)^{-1} X^{\tp} Z I_{p,q} I_{p,q} I_{p,q} Z^{\tp} X (X^{\tp}X)^{-1} I_{p,q} \\
    &= I_{p,q} (X^{\tp}X)^{-1} X^{\tp} X I_{p,q} X^{\tp} X (X^{\tp}X)^{-1} I_{p,q}
    = I_{p,q}.
  \end{align*}
\end{proof}

\ifbka
  \begin{proof}[of Proposition~\ref{identifiability}]
\else
  \begin{proof}[Proof of Proposition~\ref{identifiability}]
\fi

For $k=1,\ldots,m$, each matrix $VV^{\tp} + U_kU_k^{\tp}$ is identifiable. Moreover, since $\mathcal{G}_I$ is connected, it has no isolated nodes, implying that $\begin{bmatrix} V & U_k \end{bmatrix} $
has linearly independent columns. Thus by Lemma~\ref{lemma_indef_rotation},
\begin{equation} \label{block_rotation}
\begin{bmatrix} V & U_k \end{bmatrix} O_k = \begin{bmatrix} V' & U'_k \end{bmatrix}
\end{equation}
for $O_k$ which satisfies $S_k O_k S_k^{\tp} \in \mathcal{O}_{(p_1 + p_{2,k}),(q_1 + q_{2,k})}$ for a permutation matrix
\begin{equation*}
  S_k = \begin{bmatrix}
    I_{p_1} & 0 & 0 & 0 \\
    0 & 0 & I_{q_1} & 0 \\
    0 & I_{p_{2,k}} & 0 & 0 \\
    0 & 0 & 0 & I_{q_{2,k}} \\
\end{bmatrix}
\end{equation*}
depending on the number of assortative and disasortative latent dimensions.
It suffices to show that $O_k$ has the block structure
\begin{equation} \label{block_struct}
O_k = \begin{bmatrix} O_{11,k} & O_{12,k} \\ O_{21,k} & O_{22,k} \end{bmatrix} = \begin{bmatrix} O_{11} & 0 \\ 0 & O_{22,k} \end{bmatrix},
\end{equation}
where $O_{11} \in \mathcal{O}_{p_1,q_1}$ and $O_{22,k} \in \mathcal{O}_{p_{2,k},q_{2,k}}$.

Since $\mathcal{G}_I$ is connected, there exists a walk $\mathcal{W}$ of length $M \geq m$ on its vertices, the network layers, with vertex sequence $\{w_1, w_2, \ldots , w_M\}$ such that $\mathcal{W}$ contains each layer at least once.

Let $o_{k,\ell} \in \mathbb{R}^d$ denote the $\ell$th column of $O_k$. We will apply the following argument for each of the pairs of layers $\{w_1,w_2\},\{w_2,w_3\},,\{w_{M-1},w_M\}$ in $\mathcal{W}$.

Write
\[
o_{k,\ell} = \begin{bmatrix} o_{k,\ell}^{(1)} \\ o_{k,\ell}^{(2)} \end{bmatrix} \in \mathbb{R}^{d_1} \times \mathbb{R}^{d_{2,k}}.
\]
By \eqref{block_rotation}, for $\ell \in \{1,\ldots,d_1\}$,
\begin{equation*}
    0 = V(o_{w_1,\ell}^{(1)} - o_{w_2,\ell}^{(1)}) + U_{w_1} o_{w_1,\ell}^{(2)} - U_{w_2} o_{w_2,\ell}^{(2)}.
\end{equation*}
Then by \eqref{lin_independence},  $o_{w_1,\ell}^{(1)} - o_{w_2,\ell}^{(1)} = 0$ and $ o_{w_1,\ell}^{(2)} = o_{w_2,\ell}^{(2)} = 0.$
Since this holds for all $\ell \in \{1,\ldots,d_1\}$, we can conclude
\begin{align*}
O_{11,w_1} &= O_{11,w_2}, \\
O_{21,w_1} &= O_{21,w_2} = 0.
\end{align*}
Equating the bottom-left and bottom-right blocks of
\[
\begin{bmatrix} I_{p_1,q_1} & 0 \\ 0 & I_{p_{2,w_1},q_{2,w_1}} \end{bmatrix} = O_{w_1} \begin{bmatrix} I_{p_1,q_1} & 0 \\ 0 & I_{p_{2,w_1},q_{2,w_1}} \end{bmatrix} O_{w_1}^{\tp},
\]
we get
\begin{align*}
  O_{22,w_1} I_{p_{2,w_1},q_{2,w_1}} O_{22,w_1}^{\tp} &=  I_{p_{2,w_1},q_{2,w_1}}, \\
  O_{22,w_1}O_{12,w_1}^{\tp} &= 0.
\end{align*}
By the first equality, $O_{22,w_1} \in \mathcal{O}_{p_{2,w_1},q_{2,w_1}}$, so it is invertible. Thus the second equality gives $O_{12,w_1} = 0$. Similarly, $O_{12,w_2}=0$.

Applying this argument for each of the pairs of layers $\{w_1,w_2\},\{w_2,w_3\},\ldots,\{w_{M-1},w_M\}$ in $\mathcal{W}$ gives that $O_{11,k}$ is constant over all $k=1,\ldots,m$, and that the off-diagonal blocks are zero for each $k=1,\ldots,m$, which completes the proof by \eqref{block_struct}.
\end{proof}

\subsection{Proof of Theorem~\ref{thm1}} \label{appendix_thm1}

\ifbka
  \begin{proof}[of Theorem~\ref{thm1}]
\else
  \begin{proof}[Proof of Theorem~\ref{thm1}]
\fi

  We first outline the entire proof, which will use several technical lemmas to come later.  Let
  \begin{align*}
  \mathcal{E}_0 &= \left\{ \lVert \sum_k E_k \rVert_2 \leq 3\sigma\sqrt{mn} \right\}, \\
  \mathcal{E}_k &= \left\{ \lVert E_k \rVert_2 \leq 3\sigma \sqrt{n} \right\} \ \mathrm{for} \ k = 1, \dots, m.
                      \end{align*}
 and define the event
\begin{equation} \label{good_event}
  \mathcal{E} = \cap_{k=0}^m \mathcal{E}_k \ .
\end{equation}
We will first show in Lemma~\ref{high_prob} that $\prob(\mathcal{E}) > 1 - (m+1)n e^{-C_0 n}$ for some universal constant $C_0$.    For the remainder of the proof we assume that $\mathcal{E}$ holds.

Next,  Lemma~\ref{init_bound}  establishes the error rate \eqref{init_rate} for the initializer when $\mathcal{E}$ holds.
Then Lemmas~\ref{error_G} and \ref{error_F} show that after one iteration of proximal gradient descent, $\hat{F}^{(1)}$ and $\hat{G}_k^{(1)}$ satisfy the error bounds \eqref{thm1_bounds} when $\mathcal{E}$ holds. We also prove that when the corresponding parameters $F$ and $G_k$ have non-negative eigenvalues and $\mathcal{E}$ holds, the estimates are positive semi-definite.
Moreover, by Assumption~\ref{order_assumptions},
\begin{equation*}
    \lVert \hat{F}^{(1)} - F \rVert_F = o(n^{1/2}),
\end{equation*}
so treating $\hat{F}^{(1)}$ as the new initializer, the arguments in Lemmas~\ref{error_G} and \ref{error_F} can be repeated to establish by induction \eqref{thm1_bounds} for $\hat{F}$ and $\hat{G}_k$, the limits of the proximal gradient descent algorithm.
\end{proof}

Throughout this subsection, we let $C$ denote an arbitrary positive constant which is free of the parameters $(n,m,d_1,d_2,\sigma)$. Since the proofs of Lemmas~\ref{error_G}-\ref{error_F} only utilize the estimators after one proximal gradient step, to simplify notation we omit the superscript `$(1)$' for the estimators $\hat{F}^{(1)}$ and $\{ \hat{G}_k^{(1)} \}_{k=1}^m$.
First, we state a technical lemma from \cite{bandeira16sharp} we will use below.


\begin{lemma}[\cite{bandeira16sharp}, Corollary 3.9] \label{max_sv_band}
Let $M \in \mathbb{R}^{n \times n}$ be a symmetric matrix with entries $M_{ij} = b_{ij}g_{ij}$, where $\{g_{ij} : i \leq j\}$ are iid standard normal random variables, and \\ $\{b_{ij} : i \leq j\}$ are fixed scalars.
Define
\begin{align*}
\sigma_* &= \max_i \sqrt{ \sum_j b_{ij}^2 }, \\
\sigma_{**} &= \max_{ij} \lvert b_{ij} \rvert.
\end{align*}
Then for every $\neweps \in (0,1/2]$, there exists a constant $c_{\neweps}$ such that for every $t \geq 0$,
\begin{equation}
    \prob \left( \lVert M \rVert_2 \geq (1+\neweps)2\sigma_* + t \right) \leq n \exp \left\{ - \frac{t^2}{c_{\neweps} \sigma^2_{**}} \right\}.
\end{equation}
\end{lemma}


Using Lemma \ref{max_sv_band}, we next establish Lemma \ref{high_prob}, which shows the event $\mathcal{E}$ holds with high probability.


\begin{lemma} \label{high_prob}
Define the event $\mathcal{E}$ as in \eqref{good_event}. Then
\begin{equation*}
\prob(\mathcal{E}) \geq 1 - (m+1)n e^{- C_0 n},
\end{equation*}
for some universal constant $C_0 > 0$.
\end{lemma}

\ifbka
  \begin{proof}[of Lemma \ref{high_prob}]
\else
  \begin{proof}[Proof of Lemma \ref{high_prob}]
\fi

We first show the desired event for each individual error matrix. Fix some $k \in \{1,\ldots,m\}$. Using the notation of Proposition~\ref{max_sv_band} for the matrix $E_k$,
\[
\sigma_* = \sigma \sqrt{n-1}, \ \
\sigma_{**} = \sigma.
\]
Specify $\neweps = 1/4$ and $t = \sigma \sqrt{n} / 4$. Then by Proposition~\ref{max_sv_band},
\begin{equation*}
    \prob \left( \lVert E_k \rVert_2 \leq \frac{11}{4} \sigma \sqrt{n} \right) \geq 1 - n \exp \left\{ - C_0 n \right\}
\end{equation*}
for some constant $C_0 > 0$. In particular, $C_0$ is the universal constant $c_{1/4}$ corresponding to the choice of $\neweps = 1/4$ in Proposition~\ref{max_sv_band}.
For the matrix $\sum_k E_k$, the entries are independent with variance $m$, so
\[
\sigma_* = \sigma \sqrt{m(n-1)}, \ \
\sigma_{**} = \sigma \sqrt{m},
\]
and the same proof gives
\begin{equation*}
    \prob \left( \lVert \sum_k E_k \rVert_2 \leq 3 \sigma \sqrt{mn} \right) \geq 1 - n \exp \left\{ - C_0 n \right\} \ .
\end{equation*}
The proof is complete by a union bound.
\end{proof}


The next technical lemma will be used repeatedly to control the operator norm of a sum of square matrices.

\begin{lemma} \label{norm_sum}
Suppose $\{ M_k \}_{k=1}^m \subset \mathbb{R}^{n \times n}$ are matrices which satisfy $\max_k \lVert M_k \rVert_2 \leq \tau$, and either
\begin{equation} \label{second_order_spec}
  \max_{k_1 \neq k_2} \lVert M_{k_1}^{\tp} M_{k_2} \rVert_2 \leq \tau^2 \gamma,
\end{equation}
or
\begin{equation} \label{second_order_spec2}
  \max_{k_1 \neq k_2} \lVert M_{k_1} M_{k_2}^{\tp} \rVert_2 \leq \tau^2 \gamma.
\end{equation}
Then
\begin{equation} \label{aux_lemma1}
    \lVert \sum_k M_k \rVert_2 \leq m^{1/2} \tau (1 + m \gamma)^{1/2}.
\end{equation}
\end{lemma}

\ifbka
\begin{proof}[of Lemma~\ref{norm_sum}]
\else
\begin{proof}[Proof of Lemma~\ref{norm_sum}]
\fi

Suppose \eqref{second_order_spec} holds. Then
\begin{align*}
\lVert \sum_k M_k \rVert_2^2 &= \gamma_1\left( \left( \sum_k M_k \right)^{\tp} \left(\sum_k M_k \right) \right) = \lVert \sum_{i=1}^m \sum_{j=1}^m M_i^{\tp}M_j \rVert_2 \\
&\leq \sum_{i=1}^m \sum_{j=1}^m \lVert M_i^{\tp}M_j \rVert_2
= \sum_{i=1}^m \lVert M_i^{\tp} M_i \rVert_2 + \sum_{i=1}^m \sum_{j \neq i} \lVert M_i^{\tp}M_j \rVert_2 \\
&\leq m \tau^2 \left( 1 + m\gamma \right).
\end{align*}
If instead \eqref{second_order_spec2} holds,  the same argument can be made beginning with
\begin{equation*}
  \lVert \sum_k M_k \rVert_2^2 = \gamma_1\left( \left( \sum_k M_k \right) \left(\sum_k M_k \right)^{\tp} \right).
\end{equation*}
\end{proof}


In Lemmas~\ref{DK_for_init} and \ref{init_bound}, we establish the bound \eqref{init_rate} for the error of initializer, which will rely on an application of \cite{cai18rate}, Theorem 1.


\begin{lemma} \label{DK_for_init}
Suppose the assumptions of Theorem~\ref{thm1} hold, and suppose $\mathcal{E}$ holds.
Let $\hat{V}_0$ denote the matrix containing the first $d_1$ eigenvectors of $\frac{1}{m} \sum_k A_k$.
Then for sufficiently large $n$, 
\begin{equation}
\lVert \hat{V}_0\hat{V}_0^{\tp} - \bar{V}\bar{V}^{\tp} \rVert_2 = \lVert \sin \Theta(\bar{V},\hat{V}_0) \rVert_2 = o(d_1^{-1/2} n^{1/2 - \tau})
\end{equation}
\end{lemma}

\ifbka
\begin{proof}[of Lemma~\ref{DK_for_init}]
\else
\begin{proof}[Proof of Lemma~\ref{DK_for_init}]
\fi

We need to lower bound the smallest non-zero singular value of
\[
\bar{V}^{\tp} \left( \frac{1}{m} \sum_k A_k \right) \bar{V} \ ,
\]
which is at least
\begin{equation} \label{sv_small}
b_1n^{\tau} - \lVert \frac{1}{m} \sum_k G_k \rVert_2 - \lVert \frac{1}{m} \sum_k E_k \rVert_2 \ .
\end{equation}
Suppose $\bar{V}_{\perp}$ is an orthonormal basis for the complement of $\col(F)$. We also need to upper bound the largest singular value of
\[
\bar{V}_{\perp}^{\tp} \left( \frac{1}{m} \sum_k A_k \right) \bar{V}_{\perp},
\]
which is at most
\begin{equation} \label{sv_large}
\lVert \frac{1}{m} \sum_k G_k \rVert_2 + \lVert \frac{1}{m} \sum_k E_k \rVert_2.
\end{equation}
Then by Lemma~\ref{norm_sum},
\[
\lVert \sum_k G_k \rVert_2 \leq C m^{1/2} n^{\tau}\left( 1 + \sigma m n^{1/2 - \tau} \right)^{1/2}
\]
for a constant $C>0$. Thus the first term in \eqref{sv_large} is  $o(n^{\tau})$. Similarly, the second term is  $o(n^{\tau})$.
Finally, we need to control the operator norm $
\lVert \sum_k \bar{V}\bar{V}^{\tp}  G_k \rVert_2$.
For any $k$,
\begin{align*}
\lVert \bar{V}\bar{V}^{\tp} \bar{U}_k \Gamma_k \bar{U}_k^{\tp} \rVert_2 \leq \lVert \bar{V}^{\tp} \bar{U}_k \rVert_2 \lVert \Gamma_k \rVert_2 =  o(d_1^{-1/2} m^{1/2} n^{1/2}),
\end{align*}
and for and $k_1 \neq k_2$,
\begin{align*} \lVert \bar{V}\bar{V}^{\tp}  G_{k_1} G_{k_2}^{\tp} \bar{V}\bar{V}^{\tp} \rVert_2
  \leq  \lVert \bar{V}^{\tp} \bar{U}_{k_1} \rVert_2 \lVert \Gamma_{k_1} \rVert_2 \lVert \bar{U}_{k_1}^{\tp}\bar{U}_{k_2} \lVert \Gamma_{k_2} \rVert_2 \rVert_2 \lVert \bar{U}_{k_2}^{\tp} \bar{V} \rVert_2
= o \left( d_1^{-1}mn \cdot n^{1/2 - \tau} \right).
\end{align*}
Then by Lemma~\ref{norm_sum}, since $m n^{1/2 - \tau} \rightarrow 0$, we have
\[
\lVert \sum_k \bar{V}\bar{V}^{\tp}  G_k \rVert_2 = o(d_1^{-1/2} m n^{1/2}),
\]
and the result is a direct application of \cite{cai18rate}, Theorem 1 for sufficiently large $n$.
\end{proof}



\begin{lemma} \label{init_bound}
Denote $\Delta_{F_0} = \hat{F}^{(0)} - F$. Suppose the assumptions of Theorem~\ref{thm1} hold, and suppose $\mathcal{E}$ holds. Then for sufficiently large $n$, 
\[
\lVert \Delta_{F_0} \rVert_F = o(n^{1/2}).
\]
\end{lemma}

\ifbka
\begin{proof}[of Lemma~\ref{init_bound}]
\else
\begin{proof}[Proof of Lemma~\ref{init_bound}]
\fi

Decompose $\Delta_{F_0}$ as
\begin{align*}
\Delta_{F_0} &= \left( \hat{V}_0\hat{V}_0^{\tp} - \bar{V}\bar{V}^{\tp} \right) \left( \frac{1}{m} \sum_k A_k \right) \hat{V}_0\hat{V}_0^{\tp} + \\
&\cdots + \bar{V}\bar{V}^{\tp} \left( \frac{1}{m} \sum_k A_k \right) \left( \hat{V}_0\hat{V}_0^{\tp} - \bar{V}\bar{V}^{\tp} \right) + \\
&\cdots + \bar{V}\bar{V}^{\tp} \left( \frac{1}{m} \sum_k G_k + \frac{1}{m} \sum_k E_k \right) \bar{V}\bar{V}^{\tp}.
\end{align*}
Note that for sufficiently large $n$,
\[
\lVert \frac{1}{m} \sum_k A_k \rVert_2 \leq C n^{\tau},
\]
and by triangle inequality and Corollary~\ref{DK_for_init},
\begin{equation*}
    \lVert \Delta_{F_0} \rVert_2 = o(d_1^{-1/2}n^{1/2})
\end{equation*}
which implies
$
    \lVert \Delta_{F_0} \rVert_F = o(n^{1/2})$,
since $\rank(\Delta_{F_0}) \leq 2d_1$.
\end{proof}


With control of the error of the initializer $\hat{F}^{(0)}$, Lemma~\ref{error_G} bounds the first-iteration error for each individual matrix $G_k$ using an argument from \cite{koltchinskii11nuclear}.


\begin{lemma} \label{error_G}
For $k=1,\ldots,m$, denote $\Delta_{G_k} = \hat{G}_k - G_k$.
Suppose the assumptions of Theorem~\ref{thm1} hold, and $\mathcal{E}$ holds. Then for sufficiently large $n$, and for all $k=1,\ldots,m$,
\begin{equation}
\lVert \Delta_{G_k} \rVert_F \leq C \sigma d_2^{1/2} n^{1/2}
\end{equation}
for some constant $C>0$. Moreover, if $G_k$ has non-negative eigenvalues, $\hat{G}_k$ is positive semi-definite.
\end{lemma}

\ifbka
\begin{proof}[of Lemma~\ref{error_G}]
\else
\begin{proof}[Proof of Lemma~\ref{error_G}]
\fi

Fix $k=1,\ldots,m$. By optimality, there exists some $S_{G_k} \in \subdiff{\hat{G}_k}$ such that
\[
\iprod{ - (A_k - \hat{F}_0 - \hat{G}_k) + \lambda\alpha_k S_{G_k}}{\hat{G}_k-\tilde{G}} \leq 0
\]
where $\tilde{G}$ is some matrix with the same column space and row space as $G$. Let $\tilde{S}_G \in \subdiff{\tilde{G}}$ be arbitrary. Adding and subtracting  $\lambda\alpha_k \iprod{\tilde{S}_G}{\hat{G}_k-\tilde{G}}$ and $\iprod{F + G_k}{\hat{G}_k-\tilde{G}}$ gives
\begin{align} \label{kolt_1_G}
\iprod{\hat{F}_0-F + (\hat{G}_k - G_k)}{\hat{G}_k-\tilde{G}} + \lambda\alpha_k\iprod{S_{G_k} - \tilde{S}_G}{\hat{G}_k-\tilde{G}}
\leq \iprod{-\lambda\alpha_k\tilde{S}_G + E_k}{\hat{G}_k-\tilde{G}}.
\end{align}
By a convexity argument from \cite{koltchinskii11nuclear},
\begin{equation} \label{subdiff_arg}
\iprod{S_{G_k} - \tilde{S}_G}{\hat{G}_k-\tilde{G}} \geq 0.
\end{equation}
In particular, by convexity of the nuclear norm,
\begin{align*}
\lVert \hat{G}_k \rVert_* - \lVert \tilde{G} \rVert_* &\geq \iprod{\tilde{S}_G}{\hat{G}_k - \tilde{G}}, \\
\lVert \tilde{G} \rVert_* - \lVert \hat{G}_k \rVert_* &\geq \iprod{S_{G_k}}{\tilde{G} - \hat{G}_k},
\end{align*}
which together establish \eqref{subdiff_arg}.
Furthermore, letting
\[
\tilde{S}_G = \sum uv^{\tp} + \mathcal{P}_{G_k}^{\perp} W  \mathcal{P}_{G_k}^{\perp},
\]
for $\lVert W \rVert_2 \leq 1$, we can specify $W$ to maximize
\[
\iprod{ \mathcal{P}_{G_k}^{\perp} W  \mathcal{P}_{G_k}^{\perp}}{\hat{G}_k-\tilde{G}} = \iprod{W}{\mathcal{P}_{G_k}^{\perp} \hat{G}_k \mathcal{P}_{G_k}^{\perp}},
\]
which by duality of operator and nuclear norm gives maximum value
\[
\lVert \mathcal{P}_{G_k}^{\perp} \hat{G}_k \mathcal{P}_{G_k}^{\perp} \rVert_*.
\]
\eqref{kolt_1_G} now becomes
\begin{equation} \label{kolt_2_G}
\iprod{\hat{G}_k-G_k}{\hat{G}_k-\tilde{G}} + \lambda\alpha_k \lVert \mathcal{P}_{G_k}^{\perp} \hat{G}_k \mathcal{P}_{G_k}^{\perp} \rVert_* \leq \iprod{-\lambda\alpha_k \sum uv^{\tp} + E_k - \Delta_{F_0}}{\hat{G}_k - \tilde{G}}.
\end{equation}
We bound the first two terms of the RHS of \eqref{kolt_2_G} separately. For the first term, by duality
\begin{equation} \label{kolt_t1_G}
\lvert \iprod{-\lambda\alpha_k \sum uv^{\tp}}{\hat{G}_k - \tilde{G}}\rvert \leq \lambda\alpha_k \lVert \mathcal{P}_{G_k} (\hat{G}_k-\tilde{G}) \mathcal{P}_{G_k} \rVert_* \leq \lambda\alpha_k \sqrt{d_2} \lVert \hat{G}_k-\tilde{G} \rVert_F.
\end{equation}
For the second term,
\begin{align*}
\iprod{E_k}{\hat{G}_k-\tilde{G}} &= \iprod{E_k - \mathcal{P}_{G_k}^{\perp} E_k \mathcal{P}_{G_k}^{\perp}}{\hat{G}_k-\tilde{G}} + \iprod{\mathcal{P}_{G_k}^{\perp} E_k \mathcal{P}_{G_k}^{\perp}}{\hat{G}_k-\tilde{G}} \\
&\leq \lVert E_k - \mathcal{P}_{G_k}^{\perp} E_k \mathcal{P}_{G_k}^{\perp} \rVert_F \lVert \hat{G}_k-\tilde{G} \rVert_F + \lVert E_k \rVert_2 \lVert \mathcal{P}_{G_k}^{\perp} \hat{G}_k \mathcal{P}_{G_k}^{\perp} \rVert_*  \\
&\leq 2 \sqrt{d_2} \lVert E_k \rVert_2 \lVert \hat{G}_k-\tilde{G} \rVert_F + \lVert E_k \rVert_2 \lVert \mathcal{P}_{G_k}^{\perp} \hat{G}_k \mathcal{P}_{G_k}^{\perp} \rVert_*.
\end{align*}
Combining the previous display and \eqref{kolt_t1_G}, and specifying $\tilde{G}=G_k$, we get
\begin{align*}
\lVert \Delta_{G_k} \rVert^2 & + \left( \lambda\alpha_k - \lVert E_k \rVert_2 \right) \lVert \mathcal{P}_{G_k}^{\perp} \hat{G}_k \mathcal{P}_{G_k}^{\perp} \rVert_* \\
&\leq \left( \lambda\alpha_k \sqrt{d_2} + 2 \sqrt{d_2} \lVert E_k \rVert_2 + \lVert G_k \rVert_F \right) \lVert \Delta_{G_k} \rVert_F - \iprod{\Delta_{G_k}}{\Delta_{F_0}}.
\end{align*}
Since $\lambda\alpha_k = 3\sigma\sqrt{n} \geq \lVert E_k \rVert_2$, the second term on the left-hand side is non-negative, and the first term on the right-hand side can be bounded:
\[
\lVert \Delta_{G_k} \rVert^2 \leq C\sigma d_2^{1/2} n^{1/2} \lVert \Delta_{G_k} \rVert_F - \iprod{\Delta_{G_k}}{\Delta_{F_0}}.
\]
Dividing through by $\lVert \Delta_{G_k} \rVert_F$, we get
\[
\lVert \Delta_{G_k} \rVert_F \leq C\sigma d_2^{1/2} n^{1/2} -  \iprod{\Delta_{F_0}}{\frac{\Delta_{G_k}}{\lVert \Delta_{G_k} \rVert_F}}.
\]
Then by trace duality, and since $\lVert \Delta_{F_0} \rVert_F = o(n^{1/2})$, for sufficiently large $n$,
\[
\lVert \Delta_{G_k} \rVert_F \leq C\sigma d_2^{1/2} n^{1/2}.
\]
for some constant $C>0$, as desired.

Suppose that $G_k$ is positive semi-definite. To show $\hat{G}_k$ is positive semi-definite, fix a unit vector $v \in \mathbb{R}^n$. Note that since $\lVert \Delta_{F_0} \rVert_2 = o(n^{1/2})$, $\lVert E_k \rVert_2 + \lVert \Delta_{F_0} \rVert_2 < 3\sigma\sqrt{n} = \lambda\alpha_k$ for sufficiently large $n$. Then
\begin{align*}
  v^{\tp} \left( G_k + E_k + \Delta_{F_0} \right) v \geq  v^{\tp}(E_k + \Delta_{F_0}) v \geq - \left( \lvert E_k \rVert_2 + \lVert \Delta_{F_0} \rVert_2 \right) > - \lambda\alpha_k \ .
\end{align*}
Therefore only positive eigenvalues will survive the soft thresholding step
\begin{equation*}
  \hat{G}_k = S_{\lambda\alpha_k}(G_k + E_k - \Delta_{F_0}),
\end{equation*}
which completes the proof.
\end{proof}


In preparation to bound the norm of the sum of errors for all $G_k$ matrices, Lemmas~\ref{dk_for_G} and \ref{leq_rank} establish bounds on the recovery of their eigenvectors, and ranks.


\begin{lemma} \label{dk_for_G}
Suppose the assumptions of Theorem~\ref{thm1} hold, and $\mathcal{E}$ holds.
For $k=1,\ldots,m$, let $\hat{U}_k$ denote the first $d_2$ eigenvectors of $G_k + E_k - \Delta_{F_0}$. Then for sufficiently large $n$, and for constants $C$ and $C'$,
\begin{align} \label{dk_bound1}
\inf_{O \in \mathcal{O}_{d_2}} \lVert \hat{U}_k - \bar{U}_kO \rVert_2 &\leq C \sigma n^{1/2-\tau}, \\
\label{dk_bound2}
\lVert \hat{U}_k\hat{U}_k^{\tp} - \bar{U}_k\bar{U}_k^{\tp} \rVert_2 &\leq C' \sigma n^{1/2-\tau}.
\end{align}
\end{lemma}

\ifbka
\begin{proof}[of Lemma~\ref{dk_for_G}]
\else
\begin{proof}[Proof of Lemma~\ref{dk_for_G}]
\fi

By \cite{cai18rate}, Lemma 1 and Theorem 1, we have
\begin{equation} \label{rough_dk}
\inf_{O \in \mathcal{O}_{d_2}} \lVert \hat{U}_k - \bar{U}_kO \rVert_2 \leq \sqrt{2} \lVert \sin \Theta(\hat{U}_k,\bar{U}_k)\rVert_2 \leq C \frac{\lVert E_k - \Delta_{F_0} \rVert_2}{b_1n^{\tau}}
\end{equation}
for some constant $C$.
We also have
\begin{equation} \label{G_hat_perturb}
\lVert E_k - \Delta_{F_0} \rVert_2 \leq \lVert E_k \rVert_2 + \lVert \Delta_{F_0} \rVert_2 \leq C \sigma \sqrt{n},
\end{equation}
for some constant $C$ by Lemma~\ref{init_bound}.

Combining \eqref{G_hat_perturb} with \eqref{rough_dk} establishes \eqref{dk_bound1}. We also have by \cite{cai18rate}, Lemma 1,
\begin{equation*}
\lVert \hat{U}_k\hat{U}_k^{\tp} - \bar{U}_k\bar{U}_k^{\tp} \rVert_2 \leq 2\lVert \sin \Theta(\hat{U}_k,\bar{U}_k)\rVert_2,
\end{equation*}
which along with \cite{cai18rate}, Theorem 1, establishes \eqref{dk_bound2}.
\end{proof}


\begin{lemma} \label{leq_rank}
Suppose the assumptions of Theorem~\ref{thm1} hold, and  $\mathcal{E}$ holds.
Then for all $k=1,\ldots,m$, and for sufficiently large $n$, $\rank(\hat{G}_k) \leq d_2$.
\end{lemma}

\ifbka
\begin{proof}[of Lemma~\ref{leq_rank}]
\else
\begin{proof}[Proof of Lemma~\ref{leq_rank}]
\fi

Recall that $\hat{G}_k = S_{\lambda\alpha_k}(G_k + E_k - \Delta_{F_0})$, and $\hat{U}_k$ denotes the first $d_2$ eigenvectors of $G_k + E_k - \Delta_{F_0}$. Let $v \perp \col(\hat{U}_k)$ be an orthogonal unit vector. Then
\begin{equation}
\lVert \bar{U}_k^{\tp} v \rVert \leq \inf_{O \in \mathcal{O}_{d_2}} \lVert \bar{U}_k^{\tp}v - O\hat{U}_k^{\tp}v \rVert_2 \leq C \sigma n^{1/2 - \tau}
\end{equation}
by Lemma~\ref{dk_for_G}.  Further,
\begin{align*}
  \lvert v^{\tp}G_kv \rvert  = \lvert v^{\tp} \bar{U}_k\bar{U}_k^{\tp} G_k \bar{U}_k\bar{U}_k^{\tp} v \rvert
  \leq \lVert G_k \rVert_2 \left(  \inf_{O \in \mathcal{O}_{d_2}} \lVert \bar{U}_k - \hat{U}_kO \rVert_2 \right)^2 \ .
\leq C \sigma^2 n^{1 - \tau}
\end{align*}
Then
\begin{equation}
\lvert v^{\tp}\left(G_k + E_k - \Delta_F \right)v \rvert \leq C \sigma^2 n^{1-\tau} + \lVert E_k \rVert_2 + \lVert \Delta_{F_0} \rVert_2 \leq \frac{5}{2} \sigma\sqrt{n}
\end{equation}
for sufficiently large $n$, since $\tau > 1/2$ and $\lVert \Delta_{F_0} \rVert_2 = o(n^{1/2})$.
Then since $\lambda\alpha_k = 3\sigma\sqrt{n}$,
\[
\lvert \gamma_{d_2+1}(G_k + E_k - \Delta_{F_0}) \rvert < \lambda\alpha_k
\]
for $n$ sufficiently large, which implies $\rank(\hat{G}_k) \leq d_2$.
\end{proof}



The next technical lemma will be applied to establish the approximation error of an orthonormal basis to the direct sum of the column spaces the $G_k$ matrices. It follows from basic algebra and is given here without proof. 


\begin{lemma} \label{on_approximation}
Let $v \in \mathbb{R}^n$ be a unit vector, and $S$ be a $d$-dimensional subspace of $\mathbb{R}^n$ with orthonormal basis $U \in \mathbb{R}^{n \times d}$. Suppose $\lVert U^{\tp} v \rVert_2 \leq \neweps < 1$. Define the orthonormalization of $v$ by
\[
\tilde{v} = \frac{ \mathcal{P}_S^{\perp} v} { \lVert \mathcal{P}_S^{\perp} v \rVert_2} = \frac{ (I - UU^{\tp}) v} { \lVert (I -UU^{\tp}) v \rVert_2}.
\]
Then
\[
\lVert v - \tilde{v} \rVert_2 \leq \frac{2 \neweps}{1 - \neweps}.
\]
\end{lemma}



As further preparation to bound the norm of the sum of errors for all the $G_k$ matrices, we provide an orthonormal basis which approximates the column space of each $G_k$. Lemma~\ref{orth_basis} establishes a bound on the error of this approximation.


\begin{lemma} \label{orth_basis}
  Suppose the assumptions of Theorem~\ref{thm1} hold. Then there exists a collection mutually orthogonal matrices $\{ U_k^*\}_{k=0}^m$ such that $U_0^*$ is an orthonormal basis for
 $
  (\col(G_1) + \cdots + \col(G_m))^{\perp},
  $
  and for $k=1,\ldots,m$, $U_k^*$ satisfies
  \begin{equation*}
    \lVert \bar{U}_k - U^*_k \rVert_2 \leq C\sigma 2^{d_2/2} m^{1/2} n^{1/2 - \tau}.
  \end{equation*}
\end{lemma}

\ifbka
\begin{proof}[of Lemma~\ref{orth_basis}]
\else
\begin{proof}[Proof of Lemma~\ref{orth_basis}]
\fi

Note that the columns of each $\bar{U}_k$ form an orthonormal basis for $\col(G_k)$. Let the columns of $L$ be an orthonormal basis for
$
(\col(G_1) + \cdots + \col(G_m))^{\perp},
$
which is an $n - md_2$ dimensional subspace of $\mathbb{R}^n$. Then the columns of
\begin{equation} \label{naive_basis}
\begin{bmatrix} \bar{U}_1 & \bar{U}_2 & \cdots & \bar{U}_m & L \end{bmatrix}
\end{equation}
form a non-orthonormal basis for $\mathbb{R}^n$. Perform Gram-Schmidt orthonormalization on this matrix from left to right, which will produce a new orthonormal matrix
\begin{equation} \label{on_basis}
\begin{bmatrix} \bar{U}_1 & U^*_2 & \cdots & U^*_m & L \end{bmatrix}.
\end{equation}
Note that $U_1$ and $L$ are left unchanged. Also note that for $2 \leq k \leq m$,
\[
\col(\bar{U}_1) + \col(\bar{U}_2) + \cdots + \col(\bar{U}_k) = \col(\bar{U}_1) \oplus \col(U^*_2) \oplus \cdots \oplus \col(U^*_k).
\]
For fixed $k \in \{2,..,m\}$, and $j \in \{1,\ldots,d_2\}$, consider $u^*_{k,j}$, the $j$th column of $U^*_k$ and $\bar{u}_{k,j}$, the $j$th column of $\bar{U}_k$. By Gram-Schmidt orthonormalization,
\[
u^*_{k,j} = \frac{ \mathcal{P}^{\perp}_{S_{k,j}} \bar{u}_{k,j} }{ \lVert \mathcal{P}^{\perp}_{S_{k,j}} \bar{u}_{k,j} \rVert_2}
\]
where $S_{k,j}$ is the subspace
\[
 \col(\bar{U}_1) \oplus \col(U^*_2) \oplus \cdots \oplus \col(U^*_{k-1}) \oplus \col\left( \begin{bmatrix} u^*_{k,1} & \cdots & u^*_{k,j-1} \end{bmatrix} \right),
\]
the span of the previous orthonormal columns of \eqref{on_basis}. Note that when $j=1$, the final subspace in the direct sum is trivial.

We will use Lemma~\ref{on_approximation} repeatedly to bound $\lVert \bar{u}_{k,j} - u^*_{k,j} \rVert_2$. By construction $\bar{u}_{k,j}$ is a unit vector. An orthonormal basis for $S_{k,j}$ is given by the columns of
\[
\begin{bmatrix} \bar{U}_1 & U_2^* & \cdots & U_{k-1}^* & u^*_{k,1} & \cdots & u^*_{k,j-1} \end{bmatrix} \in \mathbb{R}^{n \times [(k-1)d_2 + (j-1)]}
\]


Then
\begin{align} \label{prev_col_orthog}
  \lVert  &   \begin{bmatrix} \bar{U}_1 & U_2^* & \cdots & U_{k-1}^* & u^*_{k,1} & \cdots & u^*_{k,j-1} \end{bmatrix}^{\tp}  \bar{u}_{k,j} \rVert_2^2
   = \lVert \begin{bmatrix} \bar{U}_1 & U_2^* & \cdots & U_{k-1}^* \end{bmatrix}^{\tp} \bar{u}_{k,j} \rVert_2^2 + \sum_{\ell=1}^{j-1} \lvert u_{k,\ell}^{*\tp} \bar{u}_{k,j} \rvert^2 \nonumber \\
&= \lVert \begin{bmatrix} \bar{U}_1 & U_2^* & \cdots & U_{k-1}^* \end{bmatrix}^{\tp} \bar{u}_{k,j} \rVert_2^2 + \sum_{\ell=1}^{j-1} \lvert (u^*_{k,\ell} - \bar{u}_{k,\ell})^{\tp} \bar{u}_{k,j} \rvert^2 \nonumber \\
&\leq \lVert \begin{bmatrix} \bar{U}_1 & U_2^* & \cdots & U_{k-1}^* \end{bmatrix}^{\tp} \bar{u}_{k,j} \rVert_2^2 + \sum_{\ell=1}^{j-1} \lVert u^*_{k,\ell} - \bar{u}_{k,\ell} \rVert^2.
\end{align}


Now \eqref{prev_col_orthog} and Lemma~\ref{on_approximation} will be applied inductively for $j=1,\ldots,d_2$.

For $j=1$, the second sum is empty and by assumption (see \eqref{eig_assump_strong}),
\[
\lVert \begin{bmatrix} \bar{U}_1 & U_2^* & \cdots & U_{k-1}^* \end{bmatrix}^{\tp}\bar{u}_{k,1} \rVert_2 \leq C \sigma (k-1)^{1/2} n^{-1/2} \leq C \sigma m^{1/2} n^{1/2 - \tau}.
\]

Then by Lemma~\ref{on_approximation},
\[
\lVert \bar{u}_{k,1} - u^*_{k,1} \rVert_2 \leq \frac{2C\sigma m^{1/2} n^{1/2 - \tau}}{1 - C\sigma m^{1/2} n^{1/2 - \tau}} \leq C\sigma m^{1/2} n^{1/2 - \tau}
\]
for sufficiently large $n$, since $mn^{1-2\tau} \rightarrow 0$.

For $j=2$, we get
\[
\lVert \begin{bmatrix} \bar{U}_1 & U_2^* & \cdots & U_{k-1}^* & u^*_{k,1} \end{bmatrix}^{\tp} \bar{u}_{k,2} \rVert_2^2 \leq 2C\sigma^2 m n^{1 - 2\tau}
\]

by \eqref{prev_col_orthog}, the orthogonality assumption, and the $j=1$ case. Then Lemma~\ref{on_approximation} gives
\[
\lVert \bar{u}_{k,2} - u^*_{k,2} \rVert_2 \leq 2C\sigma m^{1/2} n^{1/2 - \tau}
\]
for sufficiently large $n$.
Continuing for general $j \leq d_2$, the errors compound and we get
\begin{align*}
 \lVert \begin{bmatrix} \bar{U}_1 & U_2^* & \cdots & U_{k-1}^* & u^*_{k,1} & \cdots & u^*_{k,j-1} \end{bmatrix}^{\tp} \bar{u}_{k,j} \rVert_2^2   \leq C \sigma^2 \left( 1 + \sum_{\ell=1}^{j-1} 2^{\ell-1} \right) mn^{1 - 2\tau} \leq C \sigma^2 2^{d_2} mn^{1 - 2\tau} \ ,
\end{align*}

which gives
\[
\lVert \bar{u}_{k,j} - u^*_{k,j} \rVert_2 \leq C\sigma 2^{(j-1)/2} m^{1/2}n^{1/2 - \tau}
\]
for sufficiently large $n$ since $2^{d_2}mn^{1-2\tau} \rightarrow 0$.

We then use each of these to bound the squared operator norm for $k=1,\ldots,m$:
\begin{align*}
\lVert \bar{U}_k - U^*_k \rVert_2^2 & \leq \lVert \bar{U}_k - U^*_k \rVert_F^2
= \sum_{j=1}^{d_2} \lVert \bar{u}_{k,j} - u^*_{k,j} \rVert_2^2 \\
&\leq \sum_{j=1}^{d_2} C \sigma^2 2^{j-1} m n^{1 - 2\tau}  \leq C \sigma^2 2^{d_2} m n^{1 - 2\tau}.
\end{align*}
Thus, the matrix in \eqref{on_basis} provides an orthogonal decomposition of $\mathbb{R}^n$ into $m+1$ pieces, which satisfies
\begin{equation} \label{basis_approximation}
\lVert \bar{U}_k - U^*_k \rVert_2 \leq C\sigma 2^{d_2/2} m^{1/2} n^{1/2 - \tau}
\end{equation}
for $k \in \{2,\ldots,m\}$ and for some constant $C >0$. For ease of notation we will denote $U^*_1 := \bar{U}_1$, and $U^*_0 := L$, which both trivially satisfy \eqref{basis_approximation}.
\end{proof}


With Lemmas~\ref{dk_for_G}, \ref{leq_rank} and \ref{on_approximation} in hand, in Lemma~\ref{variational_bound} we establish a bound on the norm of the sum of the errors for each $G_k$ matrix.


\begin{lemma} \label{variational_bound}
Suppose the assumptions of Theorem~\ref{thm1} hold, and $\mathcal{E}$ holds.
Then for sufficiently large $n$ and some constant $C$,
\begin{equation} \label{variational_result}
\lVert \sum_{k=1}^m \Delta_{G_k} \rVert_2 \leq C \sigma m^{1/2} n^{1/2}
\end{equation}
\end{lemma}

\ifbka
\begin{proof}[of Lemma~\ref{variational_bound}]
\else
\begin{proof}[Proof of Lemma~\ref{variational_bound}]
\fi

By \eqref{eig_assump1}, for sufficiently large $n$
we have
\begin{equation} \label{G_hat_opnorm}
\lVert \hat{G}_k \rVert_2 = \lVert G_k + E_k - \Delta_{F_0} \rVert_2 - \lambda\alpha_k \leq B_1 n^{\tau} + \lVert E_k \rVert_2 + \lVert \Delta_{F_0} \rVert_2 - \lambda\alpha_k \leq Cn^{\tau}
\end{equation}
for some constant $C$, since $\tau > 1/2$.

By Lemma~\ref{leq_rank}, $\hat{G}_k$ satisfies
\[
\hat{G}_k = \hat{U}_k \hat{U}_k^{\tp} \hat{G}_k \hat{U}_k \hat{U}_k^{\tp},
\]
for sufficiently large $n$, and thus $\Delta_{G_k}$ admits the decomposition
\begin{align} \label{3_terms}
 \hat{U}_k &  \hat{U}_k^{\tp} \hat{G}_k \hat{U}_k \hat{U}_k^{\tp}   - \bar{U}_k\bar{U}_k^{\tp} G_k \bar{U}_k\bar{U}_k^{\tp} \nonumber \\
&= (\hat{U}_k \hat{U}_k^{\tp} - \bar{U}_k\bar{U}_k^{\tp} )\hat{G}_k \hat{U}_k \hat{U}_k^{\tp} + \bar{U}_k\bar{U}_k^{\tp} \Delta_{G_k} \hat{U}_k \hat{U}_k^{\tp} + \bar{U}_k\bar{U}_k^{\tp} G_k (\hat{U}_k\hat{U}_k^{\tp} - \bar{U}_k\bar{U}_k^{\tp}) \nonumber \\
&=: \termi_k + \termii_k + \termiii_k.
\end{align}
We bound the operator norm of each of these three terms separately.

{\bf Term $\termi$.}
By Lemma~\ref{dk_for_G}, \eqref{G_hat_opnorm}, and submultiplicativity,
\begin{equation}
    \lVert (\hat{U}_k \hat{U}_k^{\tp} - \bar{U}_k\bar{U}_k^{\tp} )\hat{G}_k \hat{U}_k \hat{U}_k^{\tp} \rVert_2 \leq C\sigma n^{1/2}
\end{equation}
for some constant $C>0$. We also have, for $k \neq j$ and arbitrary $O_1,O_2 \in \mathcal{O}_{d_2}$,
\begin{align*}
 \lVert \hat{U}_k^{\tp} \hat{U}_j \rVert_2 &= \lVert \hat{U}_k^{\tp} \hat{U}_j - \hat{U}_k^{\tp} \bar{U}_j O_1 +\hat{U}_k^{\tp} \bar{U}_j O_1 - O_2 \bar{U}_k^{\tp} \bar{U}_j O_1 + O_2 \bar{U}_k^{\tp} \bar{U}_j O_1 \rVert_2 \\
 &\leq \lVert \hat{U}_j - \bar{U}_j O_1 \rVert_2 + \lVert \hat{U}_k - \bar{U}_kO_2 \rVert_2 + \lVert \bar{U}_k^{\tp}\bar{U}_j \rVert_2
\end{align*}
and thus taking infimums over $\mathcal{O}_{d_2}$, and by Lemma~\ref{dk_for_G},
\begin{equation}
\lVert \hat{U}_k^{\tp} \hat{U}_j \rVert_2 \leq C \sigma n^{1/2 - \tau}
\end{equation}
for some constant $C>0$. It follows that
\begin{equation}
    \lVert (\hat{U}_k \hat{U}_k^{\tp} - \bar{U}_k\bar{U}_k^{\tp} )\hat{G}_k \hat{U}_k \hat{U}_k^{\tp} \hat{U}_j \hat{U}_j^{\tp} \hat{G}_j (\hat{U}_j \hat{U}_j^{\tp} - U_jU_j^{\tp} )\rVert_2 \leq C (\sigma n^{1/2})^2 \cdot (\sigma n^{1/2 - \tau}),
\end{equation}
so by Lemma~\ref{norm_sum},
\begin{equation} \label{term1_final}
    \lVert \sum_{k=1}^m \termi_k \rVert_2 \leq C\sigma m^{1/2} n^{1/2} \left( 1 + \sigma m n^{1/2 - \tau} \right) \leq C\sigma m^{1/2}n^{1/2}
\end{equation}
for sufficiently large $n$, by Assumption~\ref{order_assumptions}.

{\bf Term $\termii$.}  Using the variational definition of the operator norm,
\[
\lVert \sum_{k=1}^m \termii_k \rVert_2 = \sup_{x,y \in \mathbb{R}^n} \frac{ x^{\tp} \left( \sum_{k=1}^m \termii_k \right) y }{\lVert x \rVert_2 \lVert y \rVert_2}.
\]
Fix vectors $x,y \in \mathbb{R}^n$. By Lemma~\ref{orth_basis}, we can write
$
x = \sum_{i=0}^m x_i $
where
\[
x_i =  U^*_i U_i^{*\tp} x,
\] and $\{ U^*_i \}_{i=0}^m$ is a collection of mutually orthogonal matrices, with $U_0^*$ an orthonormal basis for
$
(\col(G_1) + \cdots + \col(G_m))^{\perp}$ .
For $k=1,\ldots,m$, $U_k^*$ satisfies
\begin{equation*}
  \lVert \bar{U}_k - U^*_k \rVert_2 \leq C\sigma 2^{d_2/2} m^{1/2} n^{1/2 - \tau}.
\end{equation*}
Moreover, $\sum_i \lVert x_i \rVert_2^2 = \lVert x \rVert_2^2$, and as a result,
\begin{equation} \label{decomp_sum}
\sum_i \lVert x_i \rVert_2 \leq \sqrt{2m} \lVert x \rVert_2.
\end{equation}
Decompose $y$ similarly, and write
\begin{equation} \label{terms2}
\left( \sum_{i=0}^m x_i \right)^{\tp} \left( \sum_{k=1}^m \termii_k \right) \left(\sum_{j=0}^m y_j \right) = \sum_{k=1}^m \sum_{i=0}^m \sum_{j=0}^m x_i^{\tp} \bar{U}_k\bar{U}_k^{\tp} \Delta_{G_k} \hat{U}_k \hat{U}_k^{\tp} y_j.
\end{equation}
We will bound four types of terms of \eqref{terms2} individually. First, if $i=j=k$, we have
\begin{equation}
\lvert x_k^{\tp} \bar{U}_k\bar{U}_k^{\tp} \Delta_{G_k} \hat{U}_k \hat{U}_k^{\tp} y_k \rvert \leq \lVert \Delta_{G_k} \rVert_F \lVert x_k \rVert_2 \lVert y_k \rVert_2 \leq C \sigma d_2^{1/2} n^{1/2} \lVert x_k \rVert_2 \lVert y_k \rVert_2 \ ,
\end{equation}
where the final inequality follows from Lemma~\ref{error_G}. By Cauchy-Schwarz inequality,
\[
\sum_{k=1}^m \lVert x_k \rVert_2 \lVert y_k \rVert_2 \leq \left( \sum_{k=1}^m \lVert x_k \rVert_2^2 \right)^{1/2} \left( \sum_{k=1}^m \lVert y_k \rVert_2^2 \right)^{1/2} \leq 1 \ ,
\]
so the total contribution is bounded by
\begin{equation} \label{total_t21}
C \sigma d_2^{1/2}n^{1/2} \ .
\end{equation}
If $i=k \neq j$, with $U_0$ as defined above, we have
\begin{align}
\lvert x_k^{\tp} \bar{U}_k\bar{U}_k^{\tp} \Delta_{G_k} \hat{U}_k \hat{U}_k^{\tp} y_j \rvert &= \lvert x_k^{\tp} \bar{U}_k\bar{U}_k^{\tp} \Delta_{G_k} \hat{U}_k \hat{U}_k^{\tp} U^*_j U_j^{*\tp} y_j \rvert \nonumber \\
&\leq \lVert x_k \rVert_2 \lVert \Delta_{G_k} \rVert_F \lVert \hat{U}_k U^*_j \rVert_2 \lVert y_j \rVert_2 \nonumber \\
&\leq C \sigma^2 d_2^{1/2} 2^{d_2/2} m^{1/2} n^{1 - \tau} \lVert x_k \rVert_2 \lVert y_j \rVert_2, \label{terms_22}
\end{align}
where the final inequality uses Lemma~\ref{error_G}. To bound $\lVert \hat{U}_k^{\tp} U^*_j \rVert_2$, first write
\[
\lVert \hat{U}_k^{\tp} U^*_j \rVert_2 = \lVert \hat{U}_k^{\tp} U^*_j - \hat{U}_k^{\tp} U_j + \hat{U}_k^{\tp} \bar{U}_j \rVert_2 \leq \lVert U^*_j - \bar{U}_j \rVert_2 + \lVert \hat{U}_k^{\tp}\bar{U}_j \rVert_2.
\]
By Lemma~\ref{orth_basis}, the first term is bounded by
\[
C \sigma 2^{d_2/2} m^{1/2} n^{1/2 - \tau}.
\]
For the second term, note that for any $O \in \mathcal{O}_{d_2}$,
\begin{align*}
\lVert \hat{U}_k^{\tp} \bar{U}_j \rVert_2 &= \lVert \hat{U}_k^{\tp} \bar{U}_j - O^{\tp}\bar{U}_k^{\tp}\bar{U}_j + O^{\tp}\bar{U}_k^{\tp}\bar{U}_j \rVert_2 \\
&\leq \lVert \hat{U}_k^{\tp} - (\bar{U}_kO)^{\tp} \rVert_2 \lVert \bar{U}_j \rVert_2 + \lVert O^{\tp} \rVert_2 \lVert \bar{U}_k^{\tp} \bar{U}_j \rVert_2. \\
\end{align*}
Taking an infimum over $O$ and applying Assumption~\ref{eig_assumptions}, Lemma~\ref{dk_for_G} and the fact that $\bar{U}_k^{\tp}U^*_0 = 0$ for $k=1,\ldots,m$, we get that
\begin{equation*} 
\lVert \hat{U}_k^{\tp} \bar{U}_j \rVert_2 \leq C \sigma n^{1/2 - \tau}.
\end{equation*}
Summing over the terms in \eqref{terms_22}, the total contribution is then bounded by
\begin{equation}
C \sigma^2 d_2^{1/2} 2^{d_2/2} m^{3/2} n^{1 - \tau}.
\end{equation}
If $i \neq k = j$, the total contribution can be similarly bounded. 
Finally, if $i \neq k \neq j$,
\begin{align*}
\lvert x_i^{\tp} \bar{U}_k & \bar{U}_k^{\tp} \Delta_{G_k} \hat{U}_k \hat{U}_k^{\tp} y_j \rvert = \lvert x_i^{\tp} U^*_i U^{*\tp}_i \bar{U}_k\bar{U}_k^{\tp} \Delta_{G_k} \hat{U}_k \hat{U}_k^{\tp} U^*_j U^{*\tp}_j y_j \rvert \\
                           &\leq \lVert x_i \rVert_2 \lVert U_i^{*\tp} \bar{U}_k \rVert_2 \lVert \Delta_{G_k} \rVert_F \lVert \hat{U}_k U^*_j \rVert_2 \lVert y_j \rVert_2
          \leq C \sigma^3 d_2^{1/2} 2^{d_2} m n^{3/2 - 2\tau} \lVert x_i \rVert_2 \lVert y_j \rVert_2.
\end{align*}
The total contribution can be bounded above by
\begin{equation} \label{total_t24}
C \sigma^3 d_2^{1/2} 2^{d_2} m^3 n^{3/2 - 2\tau}.
\end{equation}
Combining \eqref{total_t21}-\eqref{total_t24}, we get the upper bound
\begin{equation*}
\frac{x^{\tp} \left( \sum_{k=1}^m \termii_k \right) y}{\lVert x \rVert_2 \lVert y \rVert_2} \leq C \sigma n^{1/2} d_2^{1/2} \left( 1 + \sigma 2^{d_2/2} m^{3/2} n^{1/2 - \tau} + \sigma^2 2^{d_2} m^{3} n^{1 - 2\tau} \right),
\end{equation*}
and thus
\begin{equation} \label{term2_final}
\lVert \sum_{k=1}^m \termii_k \rVert_2 \leq C\sigma m^{1/2}n^{1/2}
\end{equation}
for sufficiently large $n$, by Assumption~\ref{order_assumptions}.

{\bf Term $\termiii$.} The proof proceeds similarly to term $\termi$, resulting in, for sufficiently large $n$,
\begin{equation} \label{term3_final}
\lVert \sum_{k=1}^m \termiii_k \rVert_2 \leq C \sigma m^{1/2}n^{1/2} \ .
\end{equation}

Combining \eqref{term1_final}, \eqref{term2_final}, and \eqref{term3_final},
\begin{equation}
\lVert \sum_{k=1}^m \Delta_{G_k} \rVert_2 \leq C\sigma m^{1/2} n^{1/2}
\end{equation}
for sufficiently large $n$.
\end{proof}


The next two lemmas apply Lemma~\ref{variational_bound} to establish that the first-iteration estimator of $F$ is low-rank. The proofs of Lemma~\ref{dk_for_F} and Lemma~\ref{leq_rank_F} take approaches analogous to the proofs of Lemma~\ref{dk_for_G} and Lemma~\ref{leq_rank} respectively, and are omitted. 


\begin{lemma} \label{dk_for_F}
Suppose the assumptions of Theorem~\ref{thm1} hold, and  $\mathcal{E}$ holds.
Let $\hat{V}$ denote the first $d_1$ eigenvectors of $F + \frac{1}{m}\sum_k E_k - \frac{1}{m} \sum_k \Delta_{G_k}$. Then for sufficiently large $n$ and  some constant $C$,
\begin{equation} \label{dk_bound1_F}
\inf_{O \in \mathcal{O}_{d_1}} \lVert \hat{V} - \bar{V}O \rVert_2 \leq C \sigma m^{-1/2} n^{1/2-\tau} \ .
\end{equation}
\end{lemma}

%
%



\begin{lemma} \label{leq_rank_F}
Suppose the assumptions of Theorem~\ref{thm1} hold, and  $\mathcal{E}$ holds. Then $\rank(\hat{F}) \leq d_1$ for sufficiently large $n$.
\end{lemma}



To complete the proof of Theorem~\ref{thm1}, Lemma~\ref{error_F} applies the same argument from \cite{koltchinskii11nuclear}, as well as Lemmas~\ref{variational_bound} and \ref{leq_rank_F} to bound the first-iteration error for $F$.


\begin{lemma} \label{error_F}
Let $\Delta_F = \hat{F} - F$. Suppose the assumptions of Theorem~\ref{thm1} hold, and $\mathcal{E}$ holds. Then for sufficiently large $n$,
\[
\lVert \Delta_F \rVert_F \leq C \sigma d_1^{1/2} m^{-1/2} n^{1/2}
\]
for some constant $C>0$. Moreover, if the eigenvalues of $F$ are non-negative, $\hat{F}$ is positive semi-definite.
\end{lemma}

\ifbka
\begin{proof}[of Lemma~\ref{error_F}]
\else
\begin{proof}[Proof of Lemma~\ref{error_F}]
\fi

By the same convexity argument as in Lemma~\ref{error_G}, we have
\begin{align*}
\lVert \Delta_F \rVert_F &\leq C \sigma d_1^{1/2} m^{-1/2} n^{1/2} - \frac{1}{m} \sum_k \iprod{\Delta_{G_k}}{\frac{\Delta_F}{\lVert \Delta_F \rVert_F}} \\
&\leq C \sigma d_1^{1/2} m^{-1/2} n^{1/2} - \frac{d_1^{1/2}}{m} \sum_k \iprod{\Delta_{G_k}}{\frac{\Delta_F}{\lVert \Delta_F \rVert_*}},
\end{align*}
where the second inequality follows by Lemma~\ref{leq_rank_F}, as long as $c_{\lambda}$ is sufficiently large.
Then by trace duality and Lemma~\ref{variational_bound},
\begin{align*}
\lVert \Delta_F \rVert_F \leq C \sigma d_1^{1/2} m^{-1/2} n^{1/2} + \frac{d_1^{1/2}}{m}\lVert \sum_k \Delta_{G_k} \rVert_2
\leq  C \sigma d_1^{1/2} m^{-1/2} n^{1/2}
\end{align*}
for sufficently large $n$, as desired. The argument that $\hat{F}$ is positive semi-definite is analogous to the argument for $\hat{G}_k$ and is omitted.
\end{proof}


\subsection{Proof of Theorem~\ref{thm2}} \label{appendix_thm2}

Structurally, the proof of Theorem~\ref{thm2} proceeds similarly to the proof of Theorem~\ref{thm1}.
The conclusion of the required auxilliary results hold under the assumptions of Theorem~\ref{thm2}, with the exception of Lemma~\ref{error_F}. In this case, we replace the application of Lemma~\ref{variational_bound} in the proof of Lemma~\ref{error_F} with the following Lemma~\ref{variational_bound2}, which gives the desired conclusion of Theorem~\ref{thm2}.

\begin{lemma} \label{variational_bound2}
  Suppose the assumptions of Theorem~\ref{thm2} hold, and  $\mathcal{E}$ holds.
  Then for sufficiently large $n$ and some constant $C$,
  \begin{equation*}
  \lVert \sum_{k=1}^m \Delta_{G_k} \rVert_2 \leq C \sigma d_2^{1/2} m^{1/2} n^{1/2} \ .
  \end{equation*}
\end{lemma}

\ifbka
\begin{proof}[of Lemma~\ref{variational_bound2}]
\else
\begin{proof}[Proof of Lemma~\ref{variational_bound2}]
\fi
  The proof proceeds similarly to the proof of Lemma~\ref{variational_bound}, beginning with the same decomposition
  \begin{equation*}
    \Delta_{G_k} = \termi_k + \termii_k + \termiii_k.
  \end{equation*}
  We can bound the contribution of $\termi$ and $\termiii$ by $Cm^{1/2}n^{1/2}$ by the same approach as Lemma~\ref{variational_bound}.
  We now bound $\termii$ using Lemma~\ref{norm_sum}, and get
  \begin{equation*}
      \lVert \sum_{k=1}^m \termii_k \rVert_2 \leq C \sigma d_2^{1/2}m^{1/2}n^{1/2}
  \end{equation*}
  for some constant $C$ and $n$ sufficiently large, completing the proof.
\end{proof}


\subsection{Proof of Proposition~\ref{prop_ase_bound}} \label{appendix_prop2}


\ifbka
\begin{proof}[of Proposition~\ref{prop_ase_bound}]
\else
\begin{proof}[Proof of Proposition~\ref{prop_ase_bound}]
\fi

We prove \eqref{ase_bound_F}; the proof of \eqref{ase_bound_G} is analogous.  Let $M^{(j)}$ denote the $j$th column of a matrix $M$.

Recall that $F = VI_{p_1,q_1}V^{\tp} = \bar{V} \Gamma_F \bar{V}^{\tp}$, and thus by Lemma~\ref{lemma_indef_rotation} $V = \bar{V} \lvert \Gamma_F \rvert^{1/2} W^{\tp}$ for some indefinite orthogonal transformation $W \in \mathcal{O}_{p_1,q_1}$.

  Suppose $n$ is sufficiently large so that the conclusions of Theorem~\ref{thm1} hold, and the event $\mathcal{E}$ holds.
  Denote the $d_1$-dimensional ASE of $\hat{F}$ by $\hat{V}$.
  For each $j=1,\ldots,d_1$, choose
  \begin{equation*}
    o^*_j = \operatorname{argmin}_{s \in \{-1,1\}} \lVert \hatbar{V}^{(j)} - s \bar{V}^{(j)} \rVert_2,
  \end{equation*}
  and denote $o^* = (o^*_1,\ldots,o^*_{d_1})^{\tp}$.  Then
  \begin{align*}
    \lVert \hat{V}^{(j)} & - o^*_j \bar{V}^{(j)} \gamma_j(F)^{1/2} \rVert_2 \\
    &= \lVert \hatbar{V}^{(j)} \gamma_j(\hat{F})^{1/2} - \hatbar{V}^{(j)} \gamma_j(F)^{1/2} + \hatbar{V}^{(j)} \gamma_j(F)^{1/2} - o^*_j\bar{V}^{(j)} \gamma_j(F)^{1/2} \rVert_2 \\
    &\leq \lvert \gamma_j(\hat{F})^{1/2} - \gamma_j(F)^{1/2} \rvert + \lvert \gamma_j(F)^{1/2} \rvert \lVert \hatbar{V}^{(j)} - o^*_j \bar{V}^{(j)} \rVert_2 \\
    &\leq \frac{\lvert \gamma_j(\hat{F}) - \gamma_j(F) \rvert}{\gamma_j(\hat{F})^{1/2} + \gamma_j(F)^{1/2}} + \lvert \gamma_j(F)^{1/2} \rvert \lVert \hatbar{V}^{(j)} - o^*_j \bar{V}^{(j)} \rVert_2 \\
    &\leq C \left( n^{-\tau/2} \left( \sigma d_1^{1/2} m^{-1/2} n^{1/2} \right) + n^{\tau/2} \left( \sigma d_1^{1/2} m^{-1/2} n^{1/2 - \xi} \right)\right) \\
    &\leq C \sigma d_1^{1/2} m^{-1/2} n^{(1+\tau)/2 - \xi}
  \end{align*}
  for sufficiently large $n$, where in the second-to-last expression, the order of the first term follows by Weyl's inequality \citep{bhatia13matrix} and Theorem~\ref{thm1}, and the order of the second term follows by \cite{cai18rate}, Corollary 1 along with Theorem~\ref{thm1} and Assumption~\ref{eig_separation}.

 Now denote
 \begin{equation*}
   W^* = I_{p_1,q_1} W I_{p_1,q_1} \operatorname{diag}(o^*) \in \mathcal{O}_{p_1,q_1},
 \end{equation*}
 where $\operatorname{diag}: \mathbb{R}^r \rightarrow \mathbb{R}^{r \times r}$ creates a diagonal matrix with the entries of the vector along the main diagonal.  Then
  \begin{align*}
    \inf_{W \in \mathcal{O}_{p_1,q_1}} \lVert \hat{V} - V W \rVert_F^2
    & \leq \lVert \hat{V} - V W^* \rVert_F^2 \leq \lVert \hat{V} - \bar{V} \lvert \Gamma_F \rvert^{1/2} \operatorname{diag}(o^*) \rVert_F^2 \\
    &= \sum_{j=1}^{d_1} \lVert \hat{V}^{(j)} - o^*_j \bar{V}^{(j)} \lvert \gamma_j(F) \rvert^{1/2} \rVert_2^2
    \leq C \sigma^2 d_1^2 m^{-1} n^{1 + \tau - 2\xi}
  \end{align*}
  which is equivalent to \eqref{ase_bound_F}.
\end{proof}

%% file: appendix_sims.tex
\ifbka
\else
  \section{Additional numerical results}
\fi


\subsection{Additional evaluation on synthetic networks} \label{appendix_sims}

We evaluate the performance of MultiNeSS in two null cases: when there is no common structure, corresponding to $d_1=0$, and when there is no individual structure, corresponding to $d_2=0$. We consider instances of the Gaussian and logistic models with no self-loops, inner product similarity, $n=400$, and $m=8$. We fix $d_2=0$ and vary $d_1 \in \{2,4,6,8,10,12\}$, and similarly fix $d_1=0$ and very $d_2 \in \{2,4,6,8,10,12\}$.
In each setting we generate 100 independent realizations of the model.
The entries of the common and individual latent position matrices are generated as independent standard normals.
We compare the MultiNeSS estimator with and without the refitting step, again denoted by MultiNeSS and MultiNeSS+, to the two non-convex oracle approaches, COSIE, and M-GRAF.
Since these cases all have either no common structure or no individual structure, we evaluate only the overall recovery of the expected value for each layer using $\mathrm{Err}_P$ as defined in Section~\ref{section_simulations}.

\begin{figure}
    \centering
    \includegraphics[width=.8\textwidth]{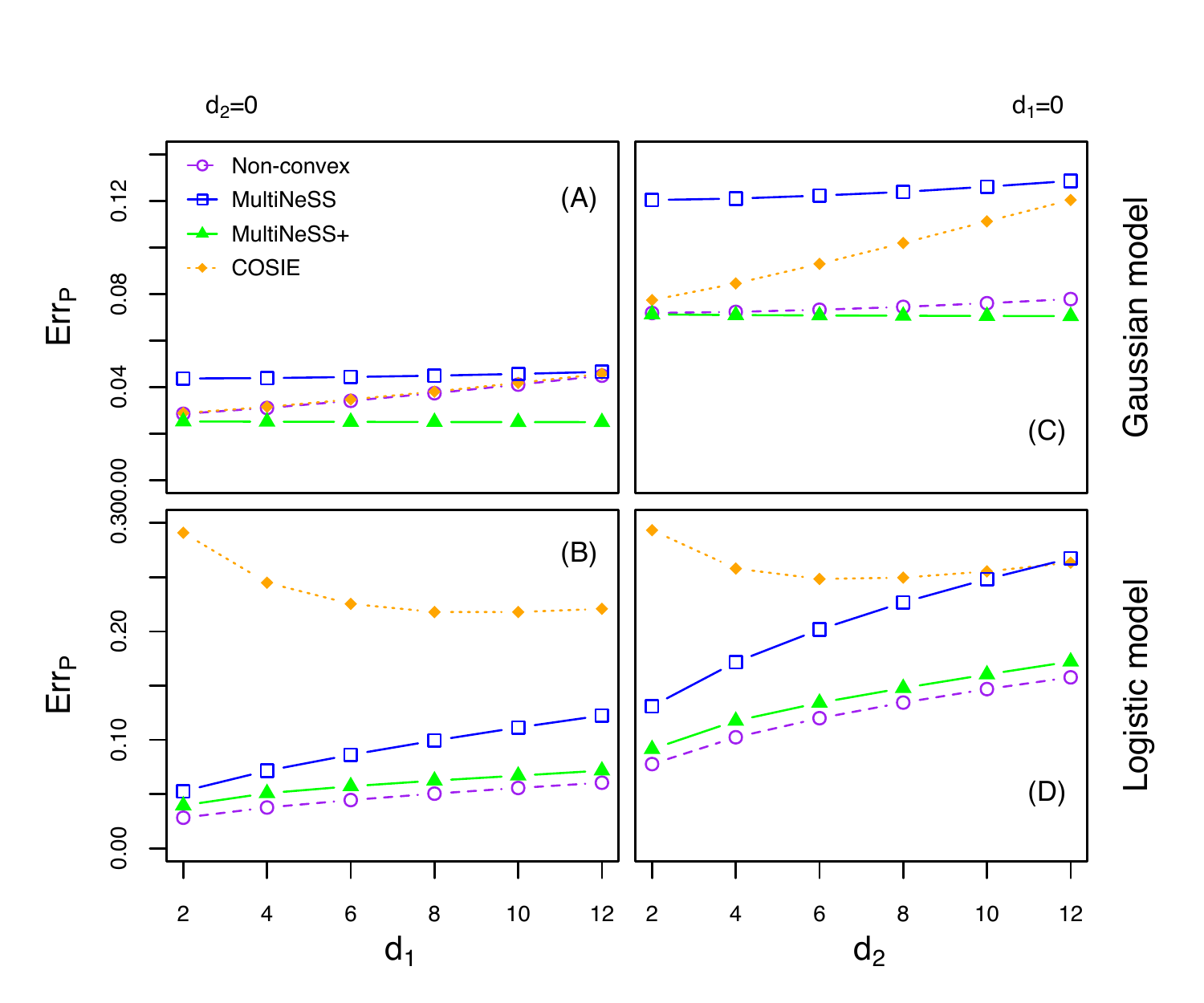}
    \caption{Frobenius norm errors for layer expectations with $d_2=0$ (left column), and $d_1=0$ (right column) for both the Gaussian model (top row) and the logistic model (bottom row).}
    \label{zerodim_plot}
\end{figure}

The results are shown in  Figure~\ref{zerodim_plot}.  Panels (A) and (B) on the left show the errors as a function of the number of common latent dimensions, with fixed $d_2=0$. Panels (C) and (D) on the right show the errors as a function of the number of individual latent dimensions, with fixed $d_1=0$.

In panels (A) and (C), errors for COSIE and the non-convex oracle increase with $d_1$ as they do not ignore the diagonal entries, which become more and more influential as $d_1$ increases. In all interations and all settings, MultiNeSS+ correctly identifies the lack of common or individual structure, returning either $\hat{d}_1 = 0$ or $\hat{d}_{2,k} = 0$ for all layers.


We also evaluate the limitations of our theoretical results by applying MultiNeSS to the Gaussian model with large $m$, and when there is non-zero correlation between the common and individual latent dimensions. We consider instances of the Gaussian model as in Section~\ref{section_gaussiansims} with $n=200$, $d_1=d_2=2$, $\sigma=1$, and $m \in \{20,50,100,150,200,300,400,600,800,1000\}$.

The entries of the common latent position matrix $V$ are independent standard normals, while the individual latent position matrices $U_k$ for $k=1,\ldots,m$ are generated as
\[
U_k = \rho V \cdot O_k + \sqrt{1-\rho^2} Z_k,
\]
where $Z_k \in \mathbb{R}^{n \times 2}$ is a matrix with independent standard normal entries, and $O_k \in \mathbb{R}^{2 \times 2}$ are random orthonormal matrices generated uniformly from $\mathcal{O}_2$.
The parameter $\rho$ controls the correlation amongst the individual latent dimensions, and between the common and individual latent dimensions.
We evaluate the performance of MultiNeSS without the refitting step for $\rho \in \{0.2,0.4,0.6,0.8\}$ using the non-normalized Frobenius norm errors for recovery of the common structure, and the overall expectation of the adjacency matrix, which we denote by $\mathrm{Err}_F^*$ and $\mathrm{Err}_P^*$ respectively.
We evaluate with non-normalized errors for this study for ease of comparison.
Since the normalizer $\lVert P_k \rVert_{\tilde{F}}^2$ for the overall structure is increasing in $\rho$, the effect of $\rho$ on recovery is much less clear if normalized errors are plotted.
Tuning parameters are fixed as $\lambda = 2.309\sqrt{nm}$ and $\alpha_k=m^{-1/2}$ for all $k=1,\ldots,m$.

\begin{figure}
    \centering
    \includegraphics[width=.8\textwidth]{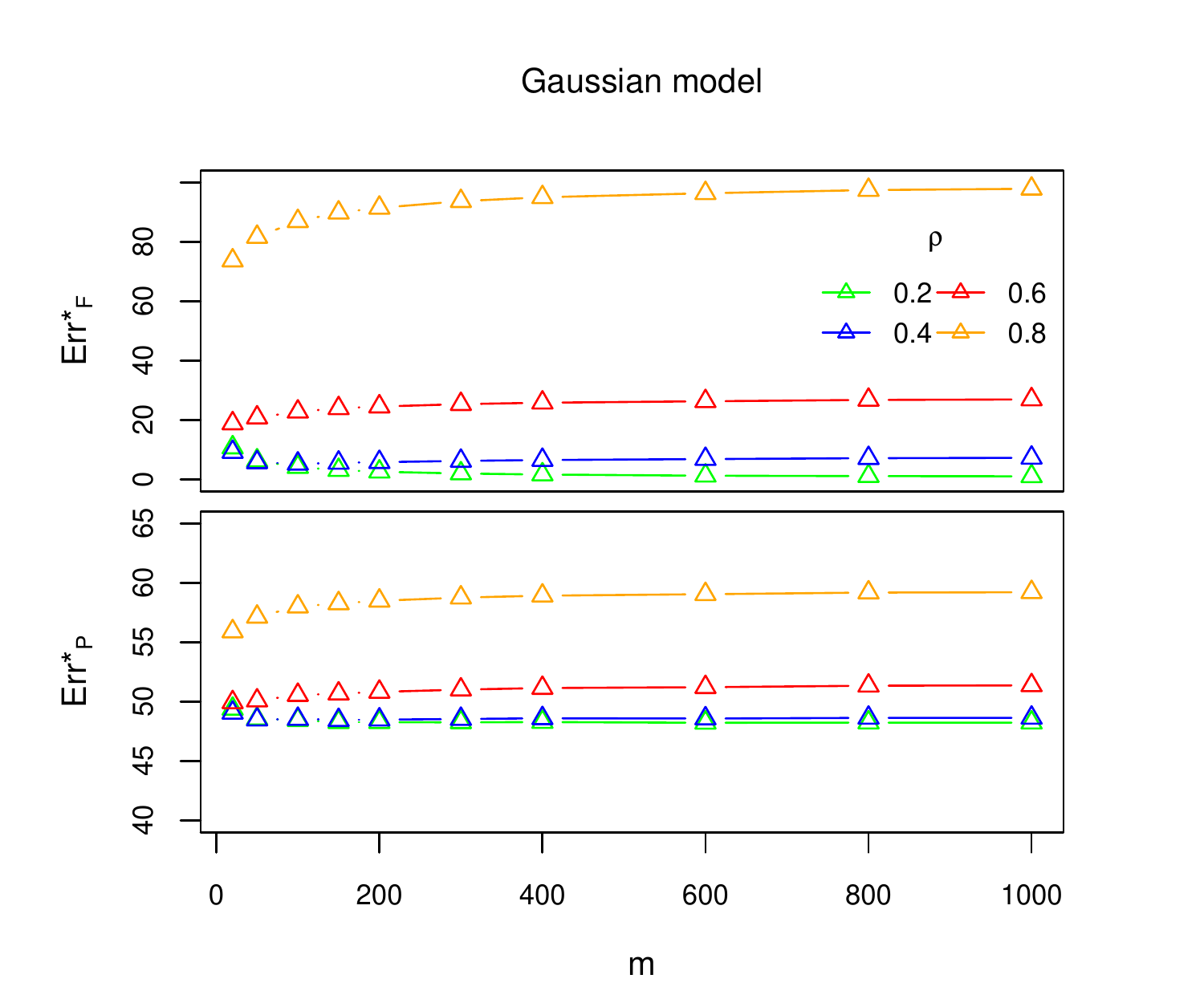}
    \caption{Frobenius norm errors for the common structure (top panel) and the overall expected value (bottom panel), under the Gaussian model.}
    \label{large_m}
\end{figure}

In the top panel of Figure~\ref{large_m}, for $\rho=0.2$, the error for $F$ decreases monotonically in $m$, while for $\rho \geq 0.6$ it increases in $m$. For $\rho=0.4$, the error decreases in $m$ for small $m$, but then reaches a minimum, and increases for larger values of $m$. 

In the bottom panel of Figure~\ref{large_m}, the error for the expected adjacency matrices is less sensitive to $\rho$, implying that the main source of error is incorrect allocation of structure between $F$ and $G_k$, not recovery of the overall structure in each layer. The performance gap in the left panel between $\rho=0.8$ and $\rho \leq 0.6$ occurs because the estimated individual matrices $G_k$ do not have the correct rank. For large $\rho$, enough of the individual structure is included in the common latent position matrix that the remaining individual structure falls below the thresholding level. Note that even for $\rho=0.2$, where the estimation error for $F$ approaches zero as $m \rightarrow \infty$, the estimation error for $P_k$ is lower bounded by the estimation error for $G_k$, which is at best constant in $m$.

\subsection{Additional trade network analysis} \label{appendix_data}

As a quantitative demonstration of our shared structure modeling approach for the multiplex trade network analyzed in Section~\ref{section_realdata}, we also apply it to an edge imputation task. For $k=1,\ldots,13$ layers, some proportion $p$ of non-zero edges in layer $k$ are held out, while the other layers are left fully observed. We compare the MultiNeSS estimator to standard low-rank matrix imputation by singular value thresholding, see for instance \cite{li20network}, denoted below by SVD, which does not incorporate information from the other layers.


\begin{figure}
    \centering
    \includegraphics[width=.8\textwidth]{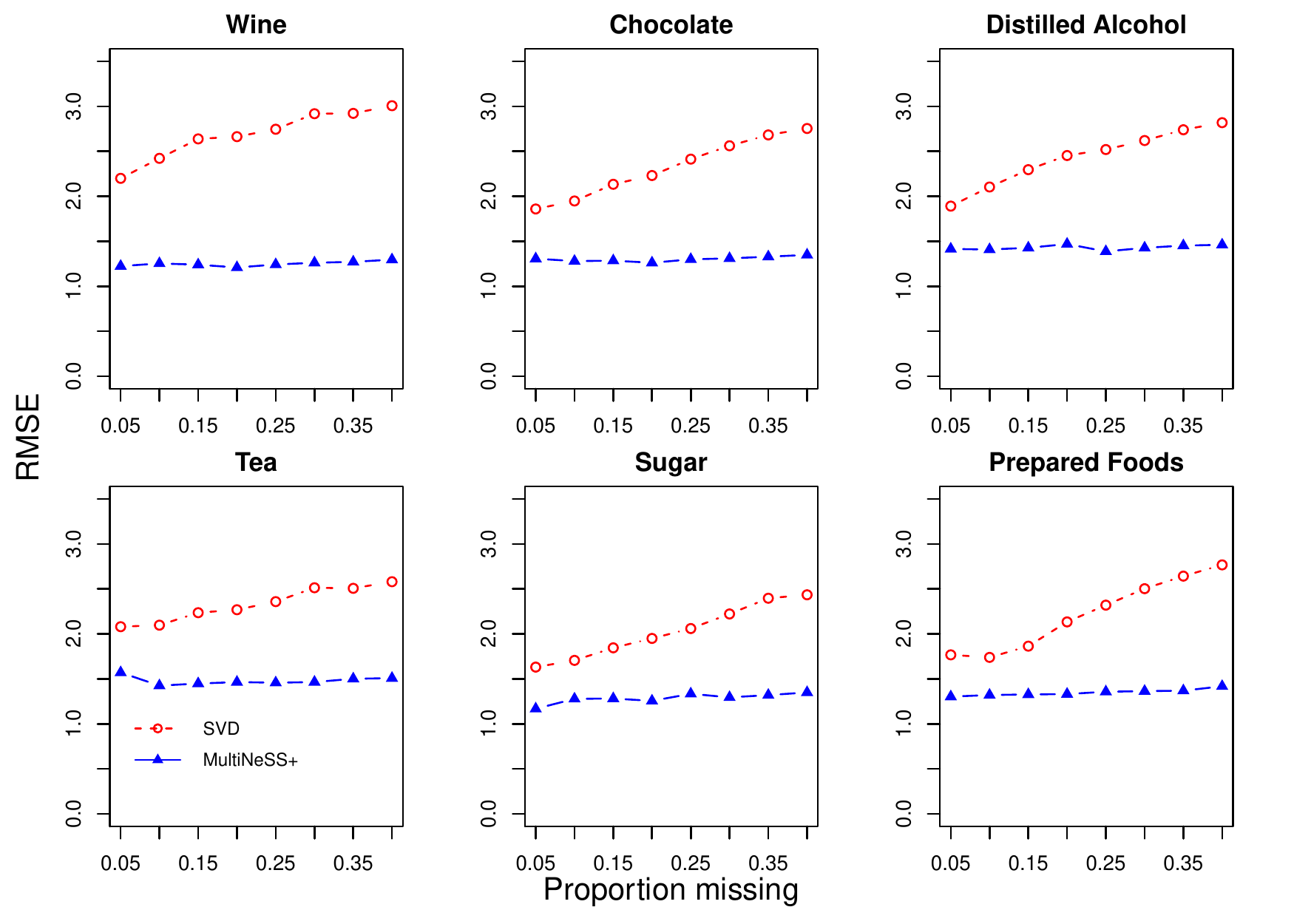}
    \caption{Comparison of edge imputation performance by root mean squared error (RMSE) for a selection of six food products.}
    \label{link_pred}
\end{figure}

From Figure~\ref{link_pred}, we see that MultiNeSS is insensitive to the proportion of edges missing compared to singular value thresholding.
In all layers MultiNeSS strictly dominates singular value thresholding, demonstrating the benefit of pooling information across layers, and validating the modeling assumption that there is common structure across all the layers of this multiplex network.